\documentclass[aps,pra,reprint,superscriptaddress,showpacs]{revtex4-1}
\usepackage[utf8]{inputenc}
\usepackage{color}
\usepackage{graphicx}
\usepackage{amssymb,amsmath,amsthm,amsfonts,mathrsfs}
\usepackage{mathtools}

\usepackage{float}
\makeatletter
\let\newfloat\newfloat@ltx
\makeatother
\usepackage{algorithm,algpseudocode}
\usepackage{xcolor}

\begin{document}
\title{Efficient motional-mode characterization for high-fidelity trapped-ion quantum computing}

\author{Mingyu Kang}
\email{mingyu.kang@duke.edu}
\affiliation{Duke Quantum Center, Duke University, Durham, NC 27701, USA}
\affiliation{Department of Physics, Duke University, Durham, NC 27708, USA}
\author{Qiyao Liang}
\affiliation{Duke Quantum Center, Duke University, Durham, NC 27701, USA}
\affiliation{Department of Physics, Duke University, Durham, NC 27708, USA}
\affiliation{IonQ, College Park, MD 20740, USA}
\author{Ming Li}
\affiliation{IonQ, College Park, MD 20740, USA}
\author{Yunseong Nam}
\affiliation{IonQ, College Park, MD 20740, USA}
\affiliation{Department of Physics, University of Maryland, College Park, MD 20742, USA}
\date{\today}

\begin{abstract}
To achieve high-fidelity operations on a large-scale quantum computer,
the parameters of the physical system must be efficiently characterized with high accuracy.
For trapped ions, the entanglement between qubits are mediated by the motional modes of the ion chain,
and thus characterizing the motional-mode parameters becomes essential.
In this paper, we develop and explore physical models that accurately predict both magnitude and sign
of the Lamb-Dicke parameters when the modes are probed {\it in parallel}.
We further devise an advanced characterization protocol that shortens the characterization time by more than an order of magnitude, when compared to that of the conventional method that only uses mode spectroscopy. We discuss potential ramifications of our results to the development of a scalable trapped-ion quantum computer, viewed through the lens of system-level resource trade offs.
\end{abstract}

\maketitle

\section{Introduction}\label{sec:Intro}

With the burgeoning of quantum computing hardware, comes the necessity to efficiently maintain and operate it. This task becomes increasingly challenging as quantum computers become larger. Indeed, in a trapped-ion quantum computer, one of the leading quantum hardware platforms today, it has been reported that keeping the fidelities of quantum gates high is already a bottleneck~\cite{hpca}. This problem is expected to get worse as the number of qubits increases.

\begin{figure*}[ht!]
\includegraphics[width=\linewidth]{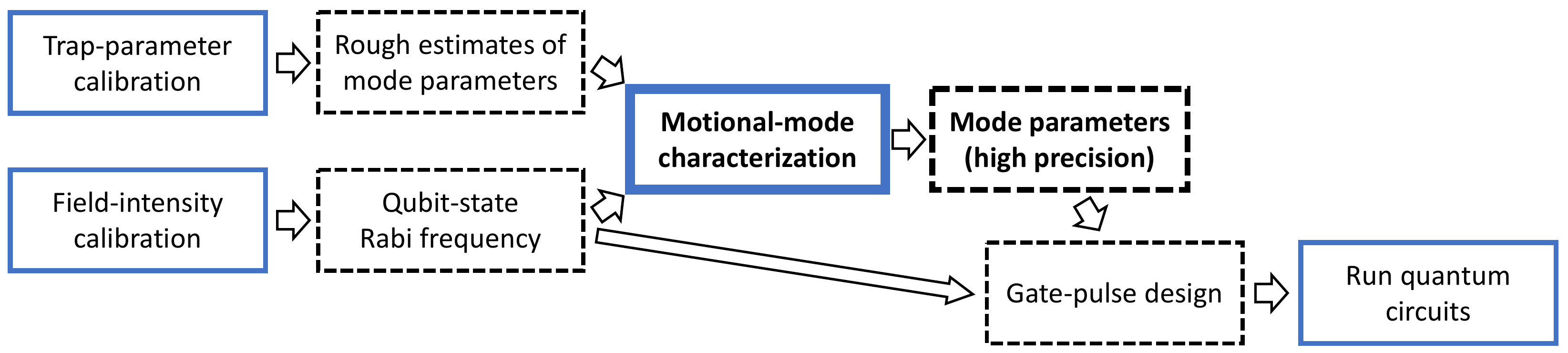}
\caption{Flowchart of a trapped-ion quantum computer's maintenance routine enabled by efficient and accurate motional-mode characterization. Each blue box with solid edges represents an experiment and each black box with dashed edges represents an obtained physical parameter. Motional-mode characterization, which measures the mode parameters with high precision, requires rough estimates of the mode parameters and precise estimates of the qubit-state Rabi frequencies, each obtained from calibration of the trap parameters [e.g., rf voltage (magnetic-field strength) and dc voltage of a Paul (Penning) trap] and the electromagnetic-field (e.g., lasers, magnetic-field gradient) intensities, respectively. Note that gate calibration, which, if performed, takes place after gate-pulse design, can be removed. This significantly reduces the time overhead of the maintenance routine, enabling more frequent and efficient updates to physical parameters that may drift over time.} 
\label{fig:intro}
\end{figure*}

Here, we too focus on improving the efficiency but from a system-parameter characterization point of view. Specifically, we aim to characterize the motional-mode parameters~\cite{Mavadia14, Goodwin16, Stutter18, Hrmo18, Welzel18, Joshi19, Hrmo19, Jarlaud20, Feng20, chen2020efficient, Sosnova21} more accurately and efficiently by leveraging parallelism. Our choice is motivated by the fact that these parameters play a crucial role in both the design and execution of entangling gate operations~\cite{PhysRevLett.74.4091, PhysRevLett.82.1835, PhysRevLett.82.1971, Blumel21, blumel2021power, li2022realizing}, one of the most apparent limiting factors for larger-scale trapped-ion quantum computing from both the fidelity and speed aspects. An efficient and accurate mode-parameter characterization can provide significant benefits, such as removing unnecessary overhead in gate calibrations that arise from incorrect parameter estimates~\cite{maksymov2021optimal, Gerster21}, enabling judicious use of hardware resources that can then be traded off for faster or more robust entangling gates~\footnote{Consider an entangling-gate pulse synthesized based on inaccurate motional-mode parameters. Even if the pulse is calibrated such that the induced gate operation is correct, the gate is suboptimal in terms of control-signal power, gate duration, and robustness, compared to a gate synthesized by a power-optimal and robust pulse-design scheme~\cite{blumel2021power, Blumel21} based on accurate motional-mode parameters.}, and opening the door to a different paradigm of quantum computer maintenance by frequent, low-cost updates to inevitably drifting parameters (See Fig.~\ref{fig:intro}). 

To this end, in our paper,
\begin{itemize}
\item We explore and exploit effective models that describe the dynamics of ions and their internal levels more accurately, thereby enabling accurate and efficient characterization,
\item We invent a more improved, faster characterization protocol than the conventional protocol that only uses mode spectroscopy, by dedicating tailor-made effort for the parameters to be estimated,
\item We leverage parallelism wherever possible to deliver efficiency, similar to the parallel gates~\cite{Grzesiak20, Bentley20} or the cooling methods~\cite{chen2020efficient} explored previously for trapped-ion quantum computers,  achieving the estimation-cost improvement from quadratic to linear in the number of qubits.
\end{itemize}

Our paper is structured as follows. In Sec.~\ref{sec:Prelim} we define the mode-characterization problem and concretely lay out our objectives. In Sec.~\ref{sec:Models} we describe in detail various models that capture different physical effects that exist in our system of interest. In Sec.~\ref{sec:Protocols} we compare and contrast the conventional and our improved experimental protocols that extract the parameters of interest. We show in Sec.~\ref{sec:Results} our results. Finally, in Sec.~\ref{sec:Discussion}, we conclude with discussions on performing mode characterization in the presence of realistic experimental sources of errors and exploiting various trade offs in system-level resources.

\section{Preliminaries}\label{sec:Prelim}

{\it Architecture --} We consider a typical trapped-ion system for universal quantum computing, where multiple ions form a Coulomb crystal and can be individually addressed by, e.g., laser beams~\cite{Wright19, Wang20, Pogorelov21}. For a quantum charge-coupled device architecture~\cite{Kielpinski02, Murali20, Pino21}, the system we consider then corresponds to a single ``gate zone.'' For a photonically interconnected architecture~\cite{Monroe14, Hucul15}, it would correspond to each ion trap containing an ion-chain.

{\it System --} Two internal states
of an atomic ion are typically used as computational qubit states.
When many ions are loaded to form a Coulomb crystal,
the external motion of the ions can be quantized and approximated
as a set of coupled quantum harmonic oscillators.
The internal and external degrees of freedom of an ion chain consisting of $N$ ions
can be described by the Hamiltonian
\begin{equation}
    \hat{H}_0 = \sum_{j=1}^{N} \frac{\omega_j^{\rm{qbt}}}{2} \hat{\sigma}_j^z + 
          \sum_{k=1}^{3N} \omega_k \hat{a}_k^{\dagger} \hat{a}_k \;,
    \label{H0}
\end{equation}
where $\omega_j^{\rm{qbt}}$ is the qubit frequency of ion $j$, 
$\omega_k$ is the normal-mode frequency of mode $k$, $\hat{\sigma}_j^z$
is the Pauli-$z$ operator in the qubit space of ion $j$,
and $\hat{a}_k$ and $\hat{a}_k^{\dagger}$ are the annihilation and creation operators for mode $k$.
We take $\hbar = 1$ throughout the manuscript.

{\it Entanglement --} A typical laser-induced multi-qubit gate operation, for instance
the M{\o}lmer-S{\o}rensen protocol~\cite{PhysRevLett.82.1835, PhysRevLett.82.1971}, 
uses the laser electric field to couple the internal and external degrees of freedom of
the participating ions. The interaction Hamiltonian $\hat{H}_{I, j}$ of a classical oscillating 
electric field of frequency $\tilde{\omega}_j$ that couples the qubit states of 
ion $j$ in an $N$-ion chain, in the rotating frame with respect to $\hat{H}_0$,
can be written as
\begin{align}\label{eq:HIj}
    \hat{H}_{I, j} &= \Omega_j 
     \hat{\sigma}^{+}_{j}
     \exp \left[-i \left( (\tilde{\omega}_j - \omega_j^{\rm{qbt}}) t + \phi_j \right) \right] \nonumber\\
     &\quad \times
    \exp \left[i\sum_{k=1}^{N'} 
    \eta_{j,k} \left( 
    \hat{a}_k e^{-i \omega_{k} t} + \hat{a}_k^\dagger e^{i \omega_{k} t}
    \right)\right]+ h.c. \;,
\end{align}
where $\hat{\sigma}^+_{j}$ is the raising operator of the $j$-th qubit, 
$\Omega_j$ is the Rabi frequency for the coupling between the
two internal states of ion qubit $j$, and $\phi_j$, hereafter chosen to be zero for brevity, is the laser phase. Also,
$\eta_{j,k}$ is the Lamb-Dicke parameter, defined as
\begin{equation}
    \eta_{j,k} = \frac{b_{j,k} |\vec{K}_{j,k}|}{\sqrt{2m \omega_k}},
\end{equation}
where $b_{j,k}$ is the $j$-th element of the normalized eigenvector of mode $k$, $m$ is the ion mass, and $\vec{K}_{j,k}$ is the wavevector of the electric field that couples ion $j$, projected along the motional direction of mode $k$.  

Typically a subset of the normal modes, indexed by $k \in \{1,2,..,N'\}$,
couples strongly when the laser field is turned on, whereas the rest of the modes
contribute negligibly.
The number of strongly coupled modes $N'$ depends on the orientation of the lasers. For example, when the laser wavevector is perfectly aligned to the trap's axial direction or one of the two radial directions, we have $N'=N$, and when it is aligned to a direction perpendicular to one of the axial or radial directions and in between the other two directions, we have $N'=2N$. A motional-mode characterization aims to estimate $\eta_{j,k}$ and $\omega_k$ of these $N'$ modes with high accuracy, as they are the ones that matter when it comes to design and implementation of high-fidelity entangling gates.

{\it Characterization --} A conventional approach for characterizing these parameters is the so-called
sideband spectroscopy, using the blue-sideband (BSB) transition. 
Experimentally, the probing lasers are set up
similarly to the ones used in multi-qubit gate operations. 
To characterize the Lamb-Dicke parameter of mode $k$ and ion $j$
and the mode frequency of mode $k$, 
we apply laser pulses of a fixed duration while
varying the laser coupling frequency $\tilde{\omega}_j$ of each pulse
near the BSB-resonance frequency $\omega_j^\mathrm{qbt}+\omega_k$. 
For each scanned frequency $\tilde{\omega}_j$,
the BSB transition near-resonantly couples 
$\left|0,n\right>_{j,k}$ and $\left|1,n+1\right>_{j,k}$, where 
$\left|a,b\right>_{j,k}$ denotes the composite state of
computational basis state $\left|a\right>_j$ of ion $j$ with $a \in \{0,1\}$
and motional Fock state $\left|b\right>_k$ of mode $k$ with phonon number $b$.
Thus, applying first a usual cooling and state-preparation procedure as the initialization step to prepare a state sufficiently close to, e.g., $\left|0,0\right>_{j,k}$,
followed by applying the aforementioned laser pulse, the population of qubit state $|1\rangle_j$ can be measured, to reveal the values of $\eta_{j,k}$ and $\omega_k$, as it
non-trivially depends on them.

Similar to other spectroscopy approaches, the conventionally used
experimental protocol is designed to probe the mode frequencies.
Note, in order to aid gate design and implementation, the Lamb-Dicke parameters also need to be accurately and efficiently characterized. The efficiency here is crucial, as there are $N \times N'$ different values of $\eta_{j,k}$ of interest.

In order to improve the efficiency, a \textit{parallelized} mode characterization is possible when the ions in a chain can be individually addressed. Specifically, all $N$ ions can be simultaneously used in probing the respectively assigned modes by setting each $\tilde{\omega}_j$ near the expected value of $\omega_k$, where mode $k$ is assigned to ion $j$.
A parallelized variant of the conventional experimental protocol descried above is 
hereafter referred to as the {\it basic protocol}, discussed in Sec.~\ref{sec:Protocols},
where an {\it improved protocol} is proposed to provide
even more efficient characterization of the Lamb-Dicke parameters.

To extract the mode parameters, one fits the measured data to a model 
that conventionally makes use of an approximated interaction Hamiltonian
\begin{align}\label{eq:baselineH}
    \hat{H}_{I, j,k} &= \Omega_j 
     \hat{\sigma}^{+}_{j}
     e^{-i \big(\tilde{\omega}_j - \omega_j^{\rm{qbt}} \big) t} \nonumber\\
     &\quad \times
    \exp \left[i
    \eta_{j,k} \left( 
    \hat{a}_k e^{-i \omega_{k} t} + \hat{a}_k^\dagger e^{i \omega_{k} t}
    \right)\right]+ h.c. \;
\end{align}
within the subspace spanned by the two states
$\left|0,0\right>_{j,k}$ and $\left|1,1\right>_{j,k}$.
We refer to such a model as the {\it baseline model},
denoted by the superscript $^{(0)}$.
The evolution operator can be written as~\cite{Wineland98} 
\begin{equation}
\label{eq:BSB}
U^{(0)}_{BSB} = \left[ \begin{matrix} 
&u_{11} \, &u_{12} \\
-&u_{12}^* \, &u_{11}^* 
\end{matrix} \right],
\end{equation}
where ${}^{*}$ denotes the complex conjugate and
\begin{align}
\label{eq:BSBele}
u_{11} &= e^{-i\Delta_{j,k} t /2} \left[\cos(X_{j,k}t) + i\frac{\Delta_{j,k}}{2X_{j,k}}\sin(X_{j,k}t)\right],\nonumber\\
u_{12} &= \frac{\Omega_{j,k}^{(0)}}{X_{j,k}} e^{-i\left(\Delta_{j,k} t /2 \right)}\sin(X_{j,k}t) \;.
\end{align}
Here, $t$ is the evolution time, 
\mbox{$\Omega^{(0)}_{j,k} = \Omega_j \eta_{j,k} e^{-\eta_{j,k}^2 / 2}$}
is the effective Rabi  frequency
between the states $\left|0,0\right>_{j,k}$ and $\left|1,1\right>_{j,k}$,
\mbox{$\Delta_{j,k} := \tilde{\omega}_j - \omega^\mathrm{qbt}_j - \omega_k$}
is the detuning from the BSB transition frequency,
and \mbox{$X_{j,k}:=([\Omega^{(0)}_{j,k}]^2 + \Delta_{j,k}^2/4)^{1/2}$}.
Inserting (\ref{eq:BSBele}) in (\ref{eq:BSB}) then applying the resulting unitary 
to the initial state $\left|0,0\right>_{j,k}$, 
we obtain the probability of ion $j$
in the bright state $|1\rangle_j$ to be~\cite{Wineland98}
\begin{equation}
\label{eq:Pt_baseline}
P^{(0)}_{j,k}(t) = 
\frac{[\Omega^{(0)}_{j,k}]^2}{[\Omega^{(0)}_{j,k}]^2 +  \frac{\Delta_{j,k}^2}{4}}
\sin^2\Bigg(\sqrt{[\Omega^{(0)}_{j,k}]^2 + \frac{\Delta_{j,k}^2}{4}} \:t\Bigg) \;,
\end{equation}
which is used to fit experimental data and extract the mode parameters.

The baseline model is approximate for two major reasons: 
(i) the spectator modes, or the modes not being probed, are ignored
and (ii) the modes are assumed to be always prepared in the motional ground state.
For a more precise estimation of the qubit population, the spectator modes' contributions, 
due to the nonzero spread of the ion's position wavepacket and the off-resonant BSB transitions,
as well as the effects of non-zero temperature, need to be taken into account. 

We emphasize that the conventional mode characterization using (\ref{eq:Pt_baseline})
does not reveal the sign of $\eta_{j,k}$ relative to one 
another~\cite{Blumel21},
which is critical for multi-qubit gate design and operation~\cite{Wu18, Grzesiak20, Bentley20, Kang21}. 
While a classical simulation of how the mode structure emerges from the ideal trapping potential can provide rough estimates of $\eta_{j,k}$'s, including their signs, it often is the case that these signs are incorrect, especially for long ion chains 
\footnote{As a concrete example, see Supplemental Material Sec. S4 of Ref.~\cite{Blumel21}. Figure S3 shows the Lamb-Dicke parameters of a seven-ion chain, measured using mode spectroscopy. The theoretical estimates of the Lamb-Dicke parameters (including their signs) are obtained by fitting the measured magnitudes of the Lamb-Dicke parameters and the mode frequencies to a theoretical model, where the fit parameters are the inter-ion spacings and the spring constants for the harmonic confinements. A naive theoretical prediction of, for example, Mode 6, from the simulated (not fitted) inter-ion spacings and spring constants, would have a symmetry with respect to the center ion, such that the center ion is a node and the left three ions have opposite sign to the right three ions. The actual measured Lamb-Dicke parameters have magnitudes and signs that are both significantly different from such predictions.}.

To guide our study of alternative characterization approaches, we summarize 
the challenges and key considerations in the following:
\begin{enumerate}
\item {\it Parallelization} -- 
There are $N \times N'$ different $\eta_{j,k}$ values of interest in an $N$-ion chain. Naively characterizing them one at a time would take $O(N^2)$ operations. To support a large-scale quantum computer, parallelization is necessary, bringing the complexity down to $O(N)$. 
\item {\it Accuracy} -- 
To characterize the mode parameters with high accuracy, 
the effect of the coupling between other modes $k' \neq k$ and qubit $j$ on the qubit population $|1\rangle_j$ needs to be taken into account. 
The coupling arises due to both the nonzero spread of the ion's position wavepacket
and the off-resonant excitation of the other modes.
\item{\it Sign problem} -- 
The relative signs of $\eta_{j,k}$ need to be distinguished,
while in (\ref{eq:Pt_baseline}) the qubit population only depends on the magnitude of $\eta_{j,k}$ and not its sign.
\item {\it Efficiency} -- 
Uncertainties in mode frequencies $\omega_k$ as well as shot noise
lead to uncertainties in estimation of $\eta_{j,k}$. 
Achieving smaller uncertainties requires considerably longer experiment time.
\end{enumerate}

These challenges lead to our objectives of efficient mode characterization: 
\begin{itemize}
    \item[] {\it Objective 1} : 
    Find effective models that better characterize the dynamics of ion qubit-state populations undergoing BSB transitions.
    \item[] {\it Objective 2} :
    Explore protocols and corresponding models that can distinguish the signs of $\eta_{j,k}$ relative to one another.
    \item[] {\it Objective 3} : 
    Find a more efficient, parallelized protocol 
    that admits minimal characterization-experiment time
    while achieving the uncertainty in estimating the mode parameters below a target value.
\end{itemize}

\section{Models}\label{sec:Models}

In this section, we detail various advanced models that predict the populations of ion qubits,
all undergoing BSB transitions in parallel. 
These models are more accurate than the traditionally used baseline model [Eq.~(\ref{eq:Pt_baseline})]
in predicting the populations and thereby characterizing the mode parameters $\eta_{j,k}$ and $\omega_k$.
In Sec.~\ref{subsec:Effects}, 
we detail three effects that occur in parallel BSB transitions that are not considered in the baseline model.
In Sec.~\ref{subsec:Models}, 
we introduce a total of five additional models, 
progressively taking the effects discussed in Sec.~\ref{subsec:Effects}, and the combinations thereof, into account,
culminating in the most sophisticated model at the end. 

\subsubsection{Effects}\label{subsec:Effects}
In this section, we discuss three effects that occur in parallel BSB transitions of ion qubits.
Considering them via more advanced models, to be detailed in the next section, leads to more accurate characterization of $\eta_{j,k}$.

\textit{(a) Non-zero temperature.}
Even after using the most sophisticated cooling techniques, 
the modes are not likely to be in the absolute motional ground state.
Therefore, the model described in Eqs.~(\ref{eq:BSB})-(\ref{eq:Pt_baseline}) is generalized to 
initial states of arbitrary phonon numbers $n$. 
The Rabi frequency for the BSB transition between $\left|0,n\right>_{j,k}$ and $\left|1,n+1\right>_{j,k}$,
assuming that states other than these two states do not affect the BSB transition, is given by
\begin{align}
\label{eq:Rabi}
\Omega^{(n)}_{j,k} &= \Omega_j 
\left|\left<n+1 \left| 
e^{i\eta_{j,k}(\hat{a}_k + \hat{a}_k^\dagger)} \right| 
n\right> \right| \nonumber\\
&= \Omega_j \frac{\eta_{j,k}}{\sqrt{n+1}} \: e^{-\eta_{j,k}^2 / 2} \: 
L_{n}^1(\eta_{j,k}^2),
\end{align}
where $L^\alpha_n$ is the generalized-Laugerre polynomial~\cite{Wineland98, Stutter18, Hrmo18}.

This generalized Rabi frequency can be used to evaluate the qubit population
undergoing the BSB transition at non-zero temperature, as we show later. 
For instance, we define $P^{(n)}_{j,k}(t)$ as the bright-state population of ion $j$
when the initial state is $\left|0,n\right>_{j,k}$, 
which is obtained by replacing $\Omega^{(0)}_{j,k}$ with $\Omega^{(n)}_{j,k}$ in (\ref{eq:Pt_baseline}).

\textit{(b) Debye-Waller (DW) effect.}
The spread of an ion's position wavepacket associated with each mode 
manifests as a reduction in the Rabi frequency, widely known as the DW effect~\cite{Wineland79}. 
In our case, even when the modes are cooled to the motional ground state,
the DW effect due to the zero-point fluctuation persists.  

When mode $k$ is being probed through ion $j$,
the DW effect due to the spectator modes $k'\neq k$ leads to 
a reduction in the Rabi frequency for the transition between $\left|0,n_k\right>_{j,k}$ and $\left|1,n_k+1\right>_{j,k}$~\cite{Stutter18, Hrmo18},
given by 
\begin{equation}
\label{eq:Rabi-DW}
\Omega_{j,k}^{(\vec{n})} = \Omega^{(n_k)}_{j,k} \prod_{k' \neq k} \bar{\mathcal{D}}_{j,k'}(n_{k'}),
\end{equation}
where $\vec{n}$ is the vector of initial phonon numbers $n_{k'}$ of mode $k'$ ($k' \in \{1,2,..,N'\}$)
and $\bar{\mathcal{D}}_{j,k'}(n_{k'})$ is the average DW reduction factor
of the spectator mode $k'$ with an initial phonon number $n_{k'}$~\cite{Wineland98}.

For an efficient characterization, each of the $N$ ions is used to probe 
the assigned mode in parallel, 
which is repeated $N'$ times with different permutations of the modes 
to probe all $N\times N'$ values of $\eta_{j,k}$.
In this case, each spectator mode $k'$ is also being probed through another ion $j'(k')$,
so its phonon number fluctuates between $n_{k'}$ and $n_{k'} + 1$.
Thus, the average DW reduction factor becomes
\begin{equation}\label{eq:avgDW-general}
    \bar{\mathcal{D}}_{j,k'}(n_{k'}) = 
    \alpha D_{j,k'}(n_{k'}) + \beta D_{j,k'}(n_{k'} + 1),
\end{equation}
where $\alpha, \beta \geq 0$ ($\alpha + \beta = 1$) are the probabilities that ion $j'(k')$ and mode $k'$ are in the state \mbox{$|0, n_{k'}\rangle_{j'(k'),k'}$} and \mbox{$|1, n_{k'}+1 \rangle_{j'(k'),k'}$}, respectively, and
\begin{align}
\label{eq:DW}
D_{j,k'}(n_{k'}) &= 
\left|\left<n_{k'} 
\left| e^{i\eta_{j,k'}(\hat{a}_{k'} + \hat{a}_{k'}^\dagger)} \right| 
n_{k'}\right> \right| \nonumber\\
&= e^{-\eta_{j,k'}^2 / 2} \: \mathscr{L}_{n_{k'}}(\eta_{j,k'}^2),
\end{align}
where $\mathscr{L}_n$ is the Laguerre polynomial.

In the case where mode $k'$ is resonantly probed for a sufficiently long evolution time, we can approximate that mode $k'$ has phonon number $n_{k'}$ half of the time and $n_{k'}+1$ for the other half. An exception is when ion $j'(k')$ is at the node of mode $k'$ ($\eta_{j'(k'), k'} \approx 0$)
and the BSB transition of ion $j'(k')$ with respect to mode $k'$ is expected to not occur.
Thus, in (\ref{eq:avgDW-general}) we apply
\begin{equation}\label{eq:alphabeta}
    (\alpha, \beta) \approx 
    \begin{cases}
    (1/2, 1/2) & \text{if}\:\eta_{j'(k'), k'} \geq \epsilon_\eta,\\
    (1, 0) & \text{if}\:\eta_{j'(k'), k'} < \epsilon_\eta,
    \end{cases}
\end{equation}
where $\epsilon_\eta$ is a discriminator that determines if $j'(k')$ is at a nodal point of mode $k'$,
typically chosen to be a small number ($\approx 10^{-4}$). 

Using (\ref{eq:Rabi-DW})-(\ref{eq:alphabeta}), Eq.~(\ref{eq:Pt_baseline}) can be further generalized
to admit non-zero initial phonon numbers of all modes,
by replacing $\Omega_{j,k}^{(0)}$ with $\Omega_{j,k}^{(\vec{n})}$.
The resulting $P_{j,k}^{(\vec{n})}(t)$ is the bright-state population of ion $j$
undergoing parallel BSB transitions, where initially all qubits are in the dark state $\left|0\right>$
and the phonon number of mode $k'$ is $n_{k'}$, the $k'$-th element of $\vec{n}$.

\textit{(c) Cross-mode coupling.}
When ion $j$ probes mode $k$, 
off-resonant BSB transitions with other modes $k' \neq k$ also occur.
The resulting effects of the other modes on the qubit state is called the cross-mode coupling. 
While the cross-mode coupling can be reduced by using a Rabi frequency $\Omega_{j,k}$ 
much smaller than the detunings $\Delta_{j,k'}$,
a smaller Rabi frequency leads to a slower BSB transition. 
Therefore, there is a trade off between reducing the error due to the cross-mode coupling and
performing a shorter characterization experiment. 

Cross-mode coupling can in principle be included in a model that simulates the evolution of the entire Hamiltonian
of $N$ ions and $N'$ modes; however, the simulation time increases exponentially with $N$.
A more realistic approach is to thus include only the nearest-neighbor modes and the ions probing them
in the simulation, limiting the simulated system size to at most three ions and three modes.

\subsubsection{Models}\label{subsec:Models}
We introduce five models of the bright-state population of the ion qubits undergoing parallel BSB transitions, improved from the baseline model in (\ref{eq:Pt_baseline}).

\textit{(a) Model 1: Debye-Waller (DW) effect.}
In our first improved model, we consider the DW effect while still assuming zero temperature.
The bright-state population $\bar{P}_{j,k}(t)$
when the initial state is $\left|0,0\right>_{j,k}$ is given by
\begin{equation}
\label{eq:Pt_DW}
\bar{P}_{j,k}(t) = P^{(\vec{0})}_{j,k}(t).
\end{equation}
Here, $P^{(\vec{0})}_{j,k}(t)$ is obtained by (\ref{eq:Pt_baseline}),
where $\Omega^{(0)}_{j,k}$ is replaced with $\Omega^{(\vec{0})}_{j,k}$,
which is found in (\ref{eq:Rabi-DW}).

Note that $\Omega^{(\vec{0})}_{j,k}$ depends not only on $\eta_{j,k}$ 
but also on other Lamb-Dicke parameters $\eta_{j,k'}$ ($k' \neq k$).
\textit{Model 1} is improved from the baseline model in that
it addresses the effects of other modes $k' \neq k$ on the population of the qubit probing mode $k$,
while taking into account that all modes are being probed in parallel. 

\textit{(b) Model 2: Non-zero temperature.}
We generalize \textit{Model 1} to include the non-zero-temperature effect. 
By admitting multiple different initial phonon numbers with the distribution function $p_{\bar{n}}(n_k)$,
where $\bar{n}$ is the average phonon number indicative of the non-zero temperature, 
we obtain the average bright-state population $\bar{P}_{j,k}(t)$ to be
\begin{equation}
\label{eq:Pt_fintemp}
    \bar{P}_{j,k}(t) = \sum_{\vec{n}} p_{\bar{n}}(\vec{n}) P^{(\vec{n})}_{j,k}(t),
\end{equation}
where $p_{\bar{n}}(\vec{n}) = \prod_k p_{\bar{n}}(n_k)$,
and $P^{(\vec{n})}_{j,k}(t)$ is found in (\ref{eq:Pt_baseline}),
where $\Omega^{(0)}_{j,k}$ is replaced with $\Omega^{(\vec{n})}_{j,k}$.
Here, for simplicity we restrict ourselves to 
thermal distributions with the same average phonon number $\bar{n}$ for all modes, 
although generalization to arbitrary distributions is straightforward. 

The summand in (\ref{eq:Pt_fintemp}) is summed over a finite number of $\vec{n}$'s 
that satisfy $p_{\bar{n}}(\vec{n}) > p_{\rm th}$ for some threshold probability $p_{\rm th}$.
In this paper we use $p_{\rm th} = 10^{-4}$ for the number of ions $N \leq 7$.
Each evaluation of $P^{(\vec{n})}_{j,k}(t)$ is parallelizable.
One can also consider sampling $\vec{n}$ randomly from the distribution $p_{\bar{n}}(\vec{n})$, 
especially for $N \gtrsim 7$ as the number of all $\vec{n}$'s to be considered becomes very large. 
In this case, the accuracy of the distribution is determined by $p_{\rm th}$,
and the sampling precision is determined by the number of samples drawn.

\textit{(c) Model 3: Time-dependent DW (TDDW) effect.}
Now we move beyond using the average DW reduction factor, 
and take into account that the reduction factor is time-dependent. 
This is because for each mode $k$ being probed through ion $j$, 
a spectator mode $k' \neq k$ is also being probed through another ion $j'(k') \neq j$,
and its phonon number fluctuates between $n_{k'}$ and $n_{k'}+1$ over time as being probed.  
The TDDW reduction factor is given by
\begin{align}
\label{eq:TDDW}
    \mathcal{D}_{j,k'}(t, n_{k'}) &= 
    \left(1 - P^{(n_{k'})}_{j'(k'),k'}(t)\right) \times D_{j,k'}(n_{k'}) \nonumber\\
    &\quad\quad + P^{(n_{k'})}_{j'(k'),k'}(t) \times D_{j,k'}(n_{k'}+1),
\end{align}
where $1 - P^{(n_{k'})}_{j'(k'),k'}(t)$ and $P^{(n_{k'})}_{j'(k'),k'}(t)$ 
are the probabilities that ion $j'(k')$ and mode $k'$ are in the state
$\left|0,n_{k'}\right>_{j'(k'),k'}$ and $\left|1,n_{k'}+1\right>_{j'(k'),k'}$ at time $t$, respectively.
Here, $P^{(n_{k'})}_{j'(k'),k'}(t)$ can be evaluated using (\ref{eq:Pt_baseline}),
where $\Omega^{(0)}_{j'(k'),k'}$ is replaced with $\Omega^{(n_{k'})}_{j'(k'),k'}$ given by (\ref{eq:Rabi}).

Now, to evaluate the population $P^{(\vec{n})}_{j,k}(t)$ with the TDDW effect considered,
the time-dependent reduction factor in (\ref{eq:TDDW}) replaces the average reduction factor that appears in (\ref{eq:Rabi-DW}),
which makes $\Omega^{(\vec{n})}_{j,k}$ time dependent as well. 
Therefore, when we numerically evaluate our model, we divide the evolution from 0 to $t$ into short time steps,
and consecutively apply the unitary evolution in (\ref{eq:BSB}), 
while updating $\Omega^{(\vec{n})}_{j,k}$ at each time step to solve for $P^{(\vec{n})}_{j,k}(t)$.
Then, we take the weighted average of the $P^{(\vec{n})}_{j,k}(t)$ over the phonon numbers $\vec{n}$ as in (\ref{eq:Pt_fintemp}) to obtain $\bar{P}_{j,k}(t)$.

\textit{(d) Model 4: Nearest neighbor (NN).}
Next, we expand the model from the one-ion, one-mode picture to the three-ion, three-mode picture
that includes the NN modes of the probed mode and their assigned qubits. 
In other words, we consider the subspace of the probed mode $k$,
its NN modes $k-1$ and $k+1$ (where the modes are ordered with increasing frequency),
and their assigned qubits $j(k)$, $j(k-1)$, and $j(k+1)$
(two-ion, two-mode for $k=1$ and $N'$). 

We use the interaction Hamiltonian
\begin{align}\label{eq:HNN}
    \hat{H}_{NN} &= i  
    \sum_{j' \in J} \Omega_{j'} \Bigg( \hat{\sigma}^{+}_{j'}
    e^{-i \left( \tilde{\omega}_{j'} - \omega_{j'}^{\rm{qbt}} \right) t}
    \prod_{k'' \notin K} \bar{\mathcal{D}}_{j',k''}(n_{k''})\nonumber \\
    &\quad \times
    \sum_{k' \in K} 
    \exp \bigg[
    i \eta_{j',k'} \bigg( 
    \hat{a}_{k'} e^{-i \omega_{k'} t} + \hat{a}^\dagger_{k'} e^{i \omega_{k'} t}
    \bigg)
    \bigg]\Bigg) \nonumber \\
    &\quad + h.c.,
\end{align}
where $J = \{ j(k-1), j(k), j(k+1) \}$, $K = \{k-1, k, k+1\}$.
The initial state is 
$|0,n_{k-1}\rangle_{j(k-1),k-1} \otimes |0,n_k\rangle_{j(k),k} \otimes |0,n_{k+1}\rangle_{j(k+1),k+1}$.
We take the matrix elements of the Hamiltonian corresponding to resonant transitions
and evaluate the unitary evolution of this Hamiltonian 
from time 0 to $t$ (see Appendix~\ref{app:simulation} for details).
The qubit population $P^{( \vec{n})}_{j(k), k}(t)$ 
is solved by projecting the state at time $t$ onto the $j(k)$-th qubit's subspace.
Finally, the average qubit population $\bar{P}_{j(k), k}(t)$ is obtained 
as in (\ref{eq:Pt_fintemp}).

Evaluating the unitary evolution of the three-ion, three-mode Hamiltonian in (\ref{eq:HNN}) takes
substantially longer time than simply evaluating trigonometric functions and polynomials as in previous models.
However, this model includes the NN modes, 
so its accuracy suffers less from the cross-mode coupling.
Note further that it properly captures the quantum interference between the qubit states and the mode states beyond a single-ion, single-mode model. 
The predicted qubit population is sensitive to the sign of $\eta_{j,k}$ relative to $\eta_{j,k \pm 1}$.

\textit{(e) Model 5: TDDW + NN.}
Finally, we combine the TDDW effect discussed in \textit{(c)} with the NN model discussed in \textit{(d)}. 
This is done by replacing the average DW reduction factor that appears in (\ref{eq:HNN}) 
with the TDDW factor in (\ref{eq:TDDW}). 

\section{Protocols} \label{sec:Protocols}

\begin{figure*}[ht!]
\includegraphics[width=0.8\linewidth]{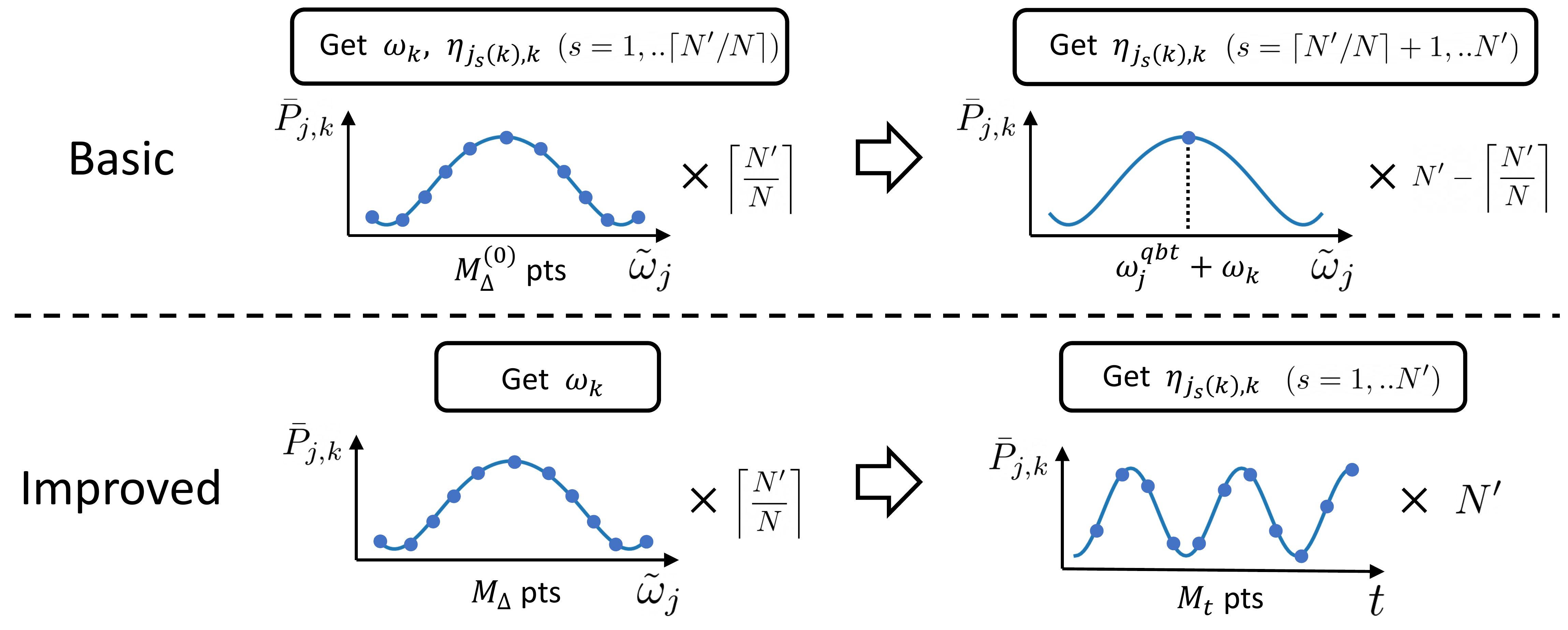}
\caption{Basic (upper) and improved (lower) protocols for mode characterization. Each experiment is performed on $N$ ions in parallel. In the basic protocol, the first step is a frequency scan, where the qubit population is measured at $M_\Delta^{(0)}$ laser coupling frequencies, repeated $\lceil N'/N \rceil$ times to measure all $N'$ values of $\omega_k$'s. In the second step, for each substep $s$ where mode $k$ is assigned to ion $j=j_s(k)$, the laser coupling frequency is fixed at $\omega_{j}^{\rm qbt} + \omega_k$, where $\omega_k$ is obtained at the first step. This is repeated for $N'-\lceil N'/N \rceil$ substeps to measure all remaining $\eta_{j,k}$'s. In the improved protocol, the first step is a similar frequency scan with $M_\Delta$ laser coupling frequencies. The second step is a time scan, where the qubit population is measured at $M_t$ evolution times, while the laser coupling frequency is again fixed at $\omega_{j}^{\rm qbt} + \omega_k$. This is repeated for $N'$ substeps to measure all $N \times N'$ values of $\eta_{j,k}$.} 
\label{fig:protocols}
\end{figure*}

In this section, we explore the effects of different experimental protocols in characterizing the mode parameters $\eta_{j,k}$ and $\omega_k$. 
The measured populations of $N$ ion qubits undergoing BSB transitions can have different sensitivities to the mode parameters for different protocols. 
Also, as there are $N \times N'$ different values of $\eta_{j,k}$, parallelization of the experimental protocol becomes a necessity. 
The conventionally used protocol is primarily designed for probing $\omega_k$.
The basic protocol, discussed in the following, is a modified version of the conventionally used protocol to probe the values of $\eta_{j,k}$ in parallel.
We then introduce an improved protocol that can more accurately and quickly determine $\eta_{j,k}$.
See Fig.~\ref{fig:protocols} for comparison.
We lastly pinpoint where the advantages of the improved protocol come from.

As will be made clear below, in order to perform efficient mode characterization, the various qubit-state Rabi frequencies and the evolution times, to be used in the forthcoming protocols, need to be chosen carefully. Considerations to be made will include: 
the expected sensitivity of the qubit populations (signals) to be measured with respect to unwanted detunings and/or cross-mode couplings and the expected qubit-population values themselves. This step requires rough estimates of the mode parameters, prior to characterizing them with high accuracy. In our experience, estimates of $\eta_{j,k}$ within an order of magnitude and those of $\omega_k$ within a few kHz tend to suffice for many of the systems used today.

\textit{Basic protocol} -- 
The basic protocol is composed of two distinct steps.
In the first step, $\omega_k$ and $\eta_{j,k}$ are measured together.
To measure all $N'$ values of $\omega_k$, this step consists of $\lceil N'/N \rceil$ repeated substeps,
where at each substep each of the $N$ ions probes its assigned mode in parallel.
Here, $\lceil \cdot \rceil$ denotes the least integer greater than or equal to the argument.
In the second step, which consists of $N' - \lceil N'/N \rceil$ substeps,
all remaining $\eta_{j,k}$'s are measured. 
The assignment of modes to ions changes at each substep.

Specifically, in the first step, one performs a \textit{frequency scan}.
At each substep, one initializes the qubit to $\left|0\right>_j$, 
excites the qubit with various laser coupling frequencies $\tilde{\omega}_j$
near the expected BSB-resonance frequency $\omega_j^\mathrm{qbt} + \omega_k$,
and measures the qubit population at a fixed time $\tau^{(0)}$.
This is performed on $N$ ions in parallel, 
such that the BSB transition on each ion $j=j_s(k)$ targets mode $k$ that is assigned at substep $s$.
Using the fact that the qubit population is maximized at zero detuning $\Delta_{j,k}=0$
when $|\Delta_{j,k}| \ll \Omega^{(0)}_{j,k}$ (which includes a sufficient range, as rough estimates of $\omega_k$ are given),
the mode frequency $\omega_k$ is measured as $\tilde{\omega}_j - \omega_j^\mathrm{qbt}$
that maximizes the population. 
The Lamb-Dicke parameter $\eta_{j,k}$ is also measured, this time 
by fitting a $\bar{P}_{j,k}(\tau^{(0)})$ expression chosen 
from any one of the models described above to the measured maximum population. 
The procedure above is repeated for $s = 1,..,\lceil N'/N \rceil$.
Note that in order to measure $\omega_k$ accurately, 
the mode assignment $j = j_s(k)$, 
the evolution time $\tau^{(0)}$, and the qubit-state Rabi frequency $\Omega_j$
all need to be prudently chosen such that the population at zero detuning is sufficiently large. 

In the second step, for each ion $j=j_s(k)$ assigned to mode $k$,
the laser coupling frequency $\tilde{\omega}_j$ is fixed at $\omega_j^\mathrm{qbt} + \omega_k$, where $\omega_k$ is known from the first step. 
The qubit population is once again measured at time $\tau^{(0)}$ and we fit a $\bar{P}_{j,k}(\tau^{(0)})$ expression chosen from any one of the models described above to the measured population, yielding $\eta_{j,k}$.
This is repeated for substeps $s=\lceil N'/N \rceil+1,..,N'$.

\textit{Improved protocol} --
The improved protocol is also composed of two steps.
In the first step, we use the frequency scan described in the basic protocol,
measuring $\omega_k$ but, importantly, not $\eta_{j,k}$.
In the second step, we perform a \textit{time scan}. 
Specifically, we fix the laser coupling frequency $\tilde{\omega}_j$ at the estimated $\omega_j^\mathrm{qbt} + \omega_k$ and 
measure the qubit population at various times \mbox{$\tau = \tau_1,.., \tau_{M_t}$} after the resonant BSB transition. 
We perform this on $N$ ions in parallel, such that each ion $j=j_s(k)$ targets the assigned mode $k$.
We repeat this for substeps $s = 1,..,N'$, exhaustively pairing $N$ ions with $N'$ modes.
Then, $\bar{P}_{j,k}(\tau_i)$ ($i = 1,..,M_t$),
selected from any one of the models we discussed previously, is fitted to each of $N \times N'$ time-scan result, in order to measure all $\eta_{j,k}$.

\begin{figure}[ht!]
\includegraphics[width=8.6cm]{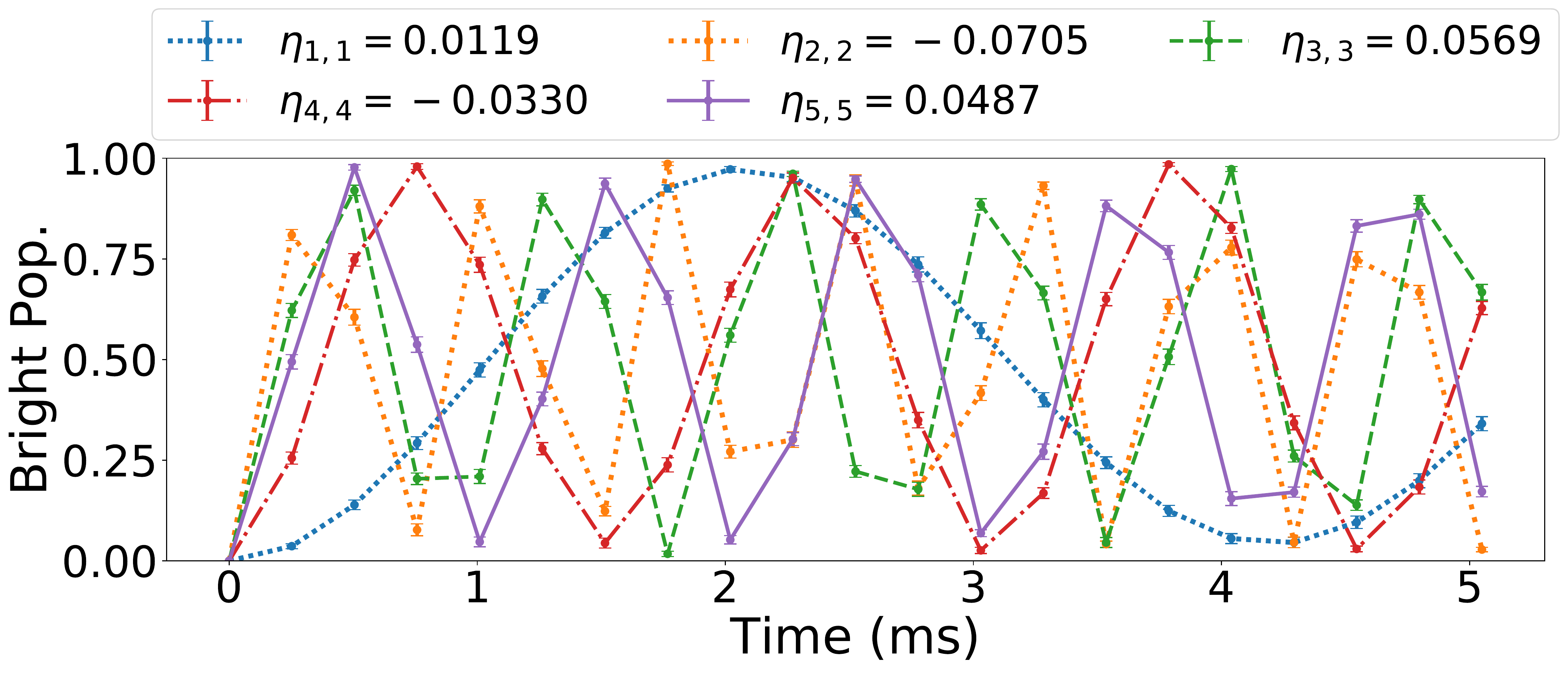}
\caption{Evolution of the bright-state populations of the qubits 
undergoing perfectly resonant ($\Delta_{j,k} = 0$) BSB transitions in parallel,
simulated using a five-ion, five-mode Hamiltonian in Eq.~(\ref{eq:HI}).
$N=N'=5$, $\Omega_j = 2\pi \times 10$ kHz $\forall j = 1,..,5$, $\bar{n} = 0.05$,
and each mode $k$ is probed through ion $j(k) = k$.
The entire set of mode parameters used here can be found in Appendix~\ref{app:values}.
The populations are recorded at $M_t=20$ equally spaced timestamps $\tau_i$.
Error bars show the shot noise for 1000 shots.}
\label{fig:timescan}
\end{figure}

Figure~\ref{fig:timescan} shows a set of example population curves that one would observe by running the time scan, to be fit using $\bar{P}_{j,k}(\tau_i)$ (not shown). The significance and difference the time scan makes for the improved protocol over the basic protocol is described next.

{\it Experimental resources} --
For each measurement, a trapped-ion quantum computer goes through a cycle of
ion-chain cooling, qubit state preparation, BSB transition, and state detection. 
For example, Ref.~\cite{hpca} shows that
the time scales of the cooling, state preparation, and state detection steps
used in a state-of-the-art trapped-ion system are of the order of
10ms, 10$\mu$s, and 100$\mu$s, respectively.
The BSB transition requires time of the order of milliseconds,
as the qubit-state Rabi frequency needs to be sufficiently small
in order to suppress the cross-mode coupling. 

For the analysis of experimental resources,
we assume $N'=N$, which corresponds to a commonly used laser-alignment setting. 
The total experiment time $T^{(0)}$ for characterizing $\eta_{j,k}$ and $\omega_k$
according to the basic protocol is then 
\begin{equation}\label{eq:exptimebl}
    T^{(0)} = M^{(0)}_\Delta S^{(0)} \tilde{\tau}^{(0)} + (N'-1) S^{(0)} \tilde{\tau}^{(0)},
\end{equation}
where $M^{(0)}_\Delta$ is the number of detunings considered in the first step, 
$S^{(0)}$ is the number of shots per data point, 
$\tilde{\tau}^{(0)}$ is the cycle time that includes the BSB-transition time $\tau^{(0)}$,
and the superscript $^{(0)}$ indicates that these values are for the basic protocol.
The counterpart for the improved protocol is
\begin{equation}\label{eq:exptimeim}
    T = M_\Delta S_\Delta \tilde{\tau}_\Delta + N' S_t \sum_{i=1}^{M_t}\tilde{\tau}_i,
\end{equation}
where $M_\Delta$ ($M_t$) is the number of detunings (timestamps) in the frequency (time) scan,
$S_\Delta$ ($S_t$) is the number of shots for each frequency (time) scan,
and $\tilde{\tau}_\Delta$ ($\tilde{\tau}_i$) is the cycle time for each frequency (time) scan 
that includes the BSB-transition time $\tau_\Delta$ ($\tau_i$).

The minimum required $T^{(0)}$ and $T$ are determined by 
the target accuracy in the $\eta_{j,k}$ estimations.
By inspecting any of the models described above, one can see that, in order to reduce uncertainties in estimating $\eta_{j,k}$, the uncertainties in $\omega_k$ need to be sufficiently small, assuming all other parameters are known. Achieving the target uncertainties in $\eta_{j,k}$ and $\omega_k$ indeed requires sufficiently large choices of $M^{(0)}_\Delta$ ($M_\Delta$), $\tau^{(0)}$ ($\tau_\Delta$), and $S^{(0)}$ ($S_\Delta$ and $S_t$) for the basic (improved) protocol.

Specific to the basic protocol, note both $\omega_k$ and $\eta_{j,k}$ simultaneously affect $\bar{P}_{j,k}(\tau^{(0)})$ directly for every data set taken according to the protocol. In other words, the population data set taken according to the basic protocol cannot distinguish the uncertainties in $\omega_k$ and $\eta_{j,k}$ separately. This results in large uncertainties in $\eta_{j,k}$ with moderate-sized uncertainties in $\omega_k$.

\begin{figure}[ht!]
\includegraphics[width=8.6cm]{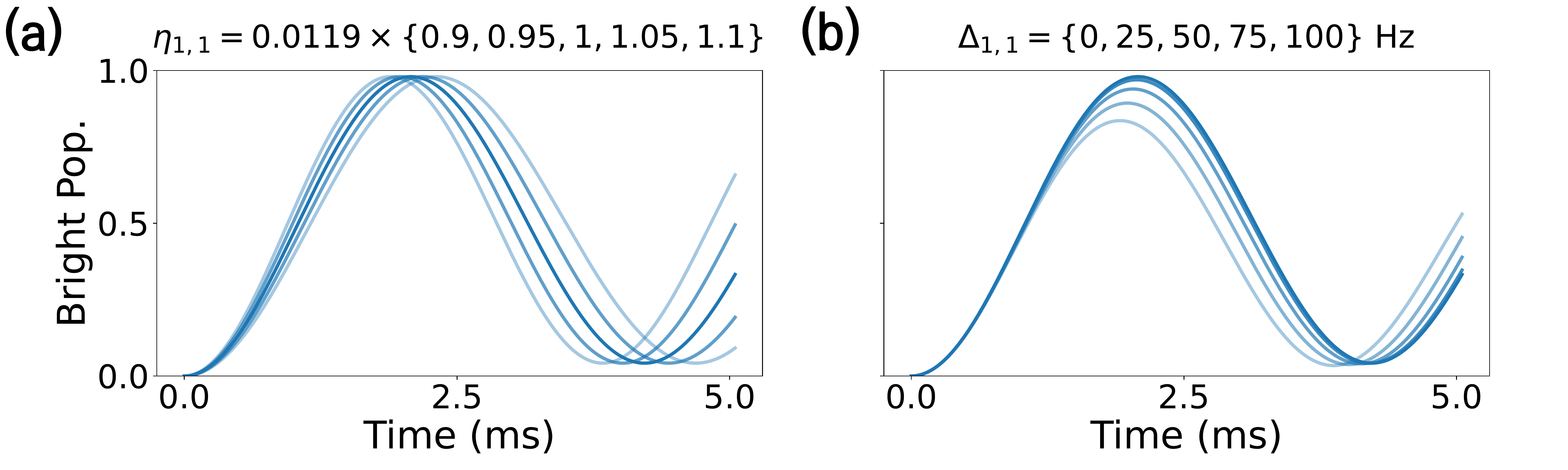}
\caption{
Bright-state population $\bar{P}_{1,1}(t)$ undergoing BSB transition as a function of time,
expected according to \mbox{\textit{Model 2}} [Eq. (\ref{eq:Pt_fintemp})],
with various values of (a) $\eta_{1,1}$ and (b) $\Delta_{1,1}$.
\mbox{$N=N'=5$}, \mbox{$\Omega_1 = 2\pi \times 10$ kHz}, and $\bar{n} = 0.05$. 
The entire set of mode parameters used here can be found in Appendix~\ref{app:values}.
The bold lines are \mbox{$\eta_{1,1} = 0.0119 \times 1$} and \mbox{$\Delta_{1,1} = 0$ Hz}, respectively. 
$\eta_{1,1}$ only affects the frequency of oscillation,
while $\Delta_{1,1}$ affects both its frequency and amplitude.
This allows for a more accurate measurement of $\eta_{j,k}$ in the presence of uncertainties in the mode frequencies.
Also note that $\bar{P}_{1,1}(t)$ is more sensitive to the value of $\eta_{1,1}$
when $\bar{P}_{1,1}(t)$ is close to 0.5, rather than close to zero or one.
}
\label{fig:signal}
\end{figure}

In contrast, for the improved protocol, uncertainties in $\eta_{j,k}$ can admit small values even when the uncertainties in $\omega_k$ are relatively large. This is achieved since,
in the improved protocol, a set of $\bar{P}_{j,k}(t)$
is measured at various $t$ values as a part of the time scan. Indeed, when fitting our models to the time-series
data, $\eta_{j,k}$ and $\Delta_{j,k}$ can be estimated
in a distinguishable way, namely, $\eta_{j,k}$ only affects 
the frequency of the oscillations of $\bar{P}_{j,k}(t)$, 
while $\Delta_{j,k}$ affects both its frequency and amplitude.
See Fig.~\ref{fig:signal} for an example.
This separation of signals for the different parameters to be estimated allows for a larger uncertainty in,
e.g., $\omega_k$ when estimating $\eta_{j,k}$ to a certain accuracy.
Targeting the same accuracy in $\eta_{j,k}$ in turn leads to
significantly shorter frequency-scan experiment time when compared to that of the basic protocol.

Figure~\ref{fig:signal}(a) shows that $\bar{P}_{j,k}(t)$ is 
maximally sensitive to the value of $\eta_{j,k}$, i.e., the rate of change in $\bar{P}_{j,k}$ with respect to change in $t$ is greatest, when $\bar{P}_{j,k}(t)$ is close to 0.5,
rather than close to zero or one. 
The improved protocol uses the entire $\bar{P}_{j,k}(t)$ curve that always includes points near 0.5.
Meanwhile, in the basic protocol where $N$ Lamb-Dicke parameters are measured in parallel,
it is challenging to find the pulse length $\tau^{(0)}$ 
such that \mbox{$\bar{P}_{j,k}(\tau^{(0)}) \approx 0.5$} for all $N$ qubits
\footnote{For the basic protocol, it is possible to tune $\Omega_j$ for each ion
such that \mbox{$\bar{P}_{j,k}(\tau^{(0)}) \approx 0.5$} for all $N$ qubits.
However, this requires percent-level prior knowledge of the values of $\eta_{j,k}$,
and therefore realistic only with, for example, an iterative protocol,
where estimates of $\eta_{j,k}$ from the previous round of characterization
are used to determine the values of $\Omega_j$ used in the next round for more accurate estimates. 
In this paper we do not provide comparison with such protocol.}.
Therefore, we expect that with the same total number of shots, 
the improved protocol leads to a smaller average uncertainty in $\eta_{j,k}$.

Fitting \textit{Models 1-5} to the experimentally measured qubit populations is a non-trivial task,
as the measured $\bar{P}_{j,k}(t)$ depends not solely on $\eta_{j,k}$,
but also on other Lamb-Dicke parameters of the spectator modes, including the nearest-neighbor modes. 
A naive approach would be to fit the model Hamiltonian of our choice to the entire set of populations $\bar{P}_{j',k'}(t)$ ($j'=1,..,N$, $k'=1,..,N'$) altogether,
where all $N \times N'$ Lamb-Dicke parameters $\eta_{j',k'}$ are fit parameters. 
However, for large $N$, determining all $N \times N'$ fit parameters at once requires too long of a
conventional-computation time for practical use.
Therefore, we employ a fitting routine that is composed of more than one iterations, where
the $\eta_{j',k'}$ ($(j',k') \neq (j,k)$) values estimated from the previous iteration
are used in the current iteration, until convergence.
The fitting routine itself can be highly parallelized so that its runtime
does not become impractically long as the number of ions $N$ increases. 
See Appendix~\ref{app:fittingroutine} for more details. 

We note in passing that while in this section
we focused on more accurate and efficient estimations of the Lamb-Dicke parameters, 
but the tool kits we provide here can indeed be readily extended for better mode-frequency estimations as well. 
For instance, fitting \textit{Models 1-5} to the qubit populations measured at various laser coupling frequencies $\tilde{\omega}_j$
can lead to more accurate estimations of $\omega_k$.

\section{Results}\label{sec:Results}

\begin{figure*}[ht!]
\includegraphics[height=4.3cm]{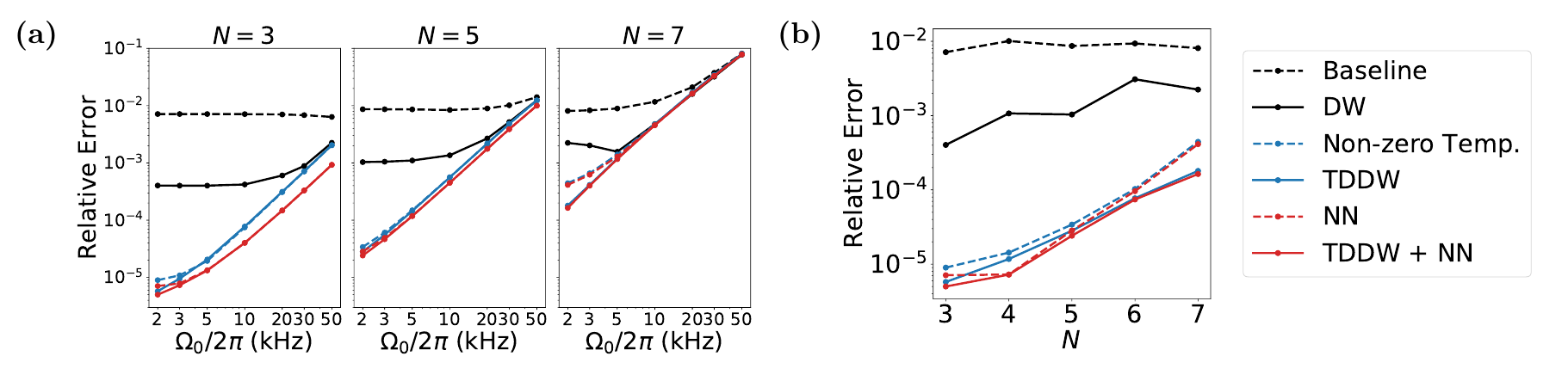}
\caption{Relative errors of $\eta_{j,k}$ for various models.
The labels are in the order of baseline and \textit{Models 1-5},
described in Sec.~\ref{sec:Models}.
Excluding the Lamb-Dicke parameters corresponding to a node \mbox{($\eta_{j,k} < 10^{-4}$)},
the errors are averaged over all $N^2$ values of $\eta_{j,k}$ for $N\leq 5$,
and averaged over $N$ values $\eta_{j=k,k}$ measured in parallel for $N>5$.
The entire set of mode parameters used here can be found in Appendix~\ref{app:values}.
(a) Relative errors with various qubit-state Rabi frequencies $\Omega_0$. 
As $\Omega_0$ decreases, error due to the cross-mode coupling decreases,
at the cost of longer experiment time. 
(b) Relative errors with various numbers of ions $N$.
$\Omega_0$ is fixed to \mbox{$2\pi \times 2$ kHz}.
}
\label{fig:errors}
\end{figure*}

In this section, we demonstrate that the three objectives of efficient mode characterization, 
stated in Sec. \ref{sec:Prelim}, can be achieved with the improved models and protocol. 
More specifically, we (i) compare the accuracy of \textit{Models 1-5} to the baseline model
in measuring the Lamb-Dicke parameters $\eta_{j,k}$, 
(ii) demonstrate that \textit{Model 4} can distinguish the relative signs of $\eta_{j,k}$,
and (iii) show that the improved protocol leads to significantly shorter characterization-experiment time
than the basic protocol for a given target accuracy in $\eta_{j,k}$ estimations. 

To perform numerical tests, we simulate the parallel BSB-transition experiment.
The BSB Hamiltonian in the interaction picture is given by
\begin{equation} \label{eq:HI}
    \hat{H}_{I} = \sum_{j=1}^N \hat{H}_{I,j},
\end{equation}
where $\hat{H}_{I,j}$ is found in (\ref{eq:HIj}).
To enable efficient simulations, we apply approximations detailed in Appendix~\ref{app:simulation}.
Also, in this section we assume $N' = N$, which agrees with a typical laser alignment. 
The evolution operator implied by $\hat{H}_{I}$ is applied to all initial states 
$\bigotimes_{k'=1}^{N} \left|0, n_{k'}\right>_{j'(k'), k'}$,
where ion $j'(k')$ is the ion assigned for mode $k'$,
and the vector of phonon numbers $\vec{n}$ satisfies \mbox{$p_{\bar{n}}(\vec{n}) > p_{\rm th}$}, as discussed previously.
We use average phonon number $\bar{n} = 0.05$ for all modes and $p_{\rm th} = 10^{-4}$, as a concrete example that is consistent with the state-of-the-art experiments today.
The state at time $t$ is projected onto the qubits' subspace and yields the qubit populations $P^{(\vec{n})}_{j(k),k}(t)$ for all $k$. 
Finally, we compute the weighted average $\bar{P}_{j(k),k}(t)$ as in (\ref{eq:Pt_fintemp}),
which are then used in the fitting procedure for the previously discussed models
to test the respective models' accuracy.
Note the Hilbert space dimension grows exponentially with the number of ions $N$.
We thus test our models up to $N = 7$, keeping the simulation time reasonable.

\subsection{Accuracy}
\label{subsec:Accuracy}

We compare the baseline model and $\textit{Models 1-5}$ in their performance in adequately capturing 
the qubit-population evolution obtained from our simulation.
Here, as an example we assume that all ions are simultaneously driven with the same qubit-state Rabi frequency:
\mbox{$\Omega_j = \Omega_0$ $\forall j = 1,..,N$}.
The populations are recorded at \mbox{$M_t=20$} equally spaced timestamps $\tau_i$.
The longest timestamp is chosen as 
\mbox{$\tau_{M_t} = 2.5 \sqrt{N} (\Omega_0 |\langle \vec{K} \rangle| /\sqrt{2m \langle \omega_{\rm mode} \rangle})^{-1}$},
where $\langle \vec{K} \rangle$ is a rough estimate of the projected wavevector and $\langle \omega_{\rm mode} \rangle$ is a rough estimate of the average of mode frequencies, 
such that the longest BSB transition with respect to the center-of-mass mode
undergoes roughly five Rabi half-cycles for all $N$ and $\Omega_0$. 

Figure~\ref{fig:errors} shows the mean relative errors in estimating $\eta_{j,k}$,
obtained from using various models, as a function of $\Omega_0$ and $N$. 
Here the relative error is defined as \mbox{$|(\eta_{j,k} - \eta^{\rm (est)}_{j,k}) / \eta_{j,k}|$},
where $\eta^{\rm (est)}_{j,k}$ is the estimated Lamb-Dicke parameter from fitting.
In general, \textit{Models 1-5} show significant improvement in the estimation accuracy of $\eta_{j,k}$ compared to the baseline model. 
In particular, the relative error of size less than $10^{-3}$ can only be achieved by using the improved models. 
Including both the DW effect from the spectator modes and the non-zero temperature effect
significantly reduces the error,
especially when these effects are larger than the effect of the cross-mode coupling,
which occurs when $\Omega_0$ is small. 

Figure~\ref{fig:errors}(a) shows that \textit{Models 2-5} exhibit a power-law behavior,
relative error being proportional to $\Omega_0^2$.
Note we are in the perturbative regime 
where the Rabi frequency $\Omega_{j,k} \propto \Omega_0$ is much smaller than
the detuning $\Delta_{j,k'}$ from modes $k' \neq k$ not being probed by ion $j$. 
The observed power law (linear trend in the log-log plot) is reminiscent of the dominance of the cross-mode-coupling error
in this regime of $\Omega_0$, absent other sources of dominant uncertainties, mentioned and taken care of previously. 

One would expect that including the NN modes in the model reduces the error from the cross-mode coupling.
Indeed, \textit{Model 4} and \textit{5} have noticeably smaller errors than \textit{Model 2} and \textit{3} for $N < 5$.
However, for longer ion chains, the errors do not have as much difference. 
In the case where, for example, $\eta_{j, k \pm 1}$ are smaller than $\eta_{j, k \pm 2}$,
the effects of the modes $k \pm 2$ can be comparable to or larger than those of the NN modes $k \pm 1$
on the error in measuring $\eta_{j,k}$.
For such cases, the NN model can be extended to include the modes with significant effects,
at the cost of longer computation time for fitting. 

The models with the TDDW effect included achieve the highest accuracy.
For instance, in Fig.~\ref{fig:errors}(b), when $N=7$, 
the errors of \textit{Models 3} and \textit{5} 
are 2.5 times smaller than those of \textit{Models 2} and \textit{4}.
Based on the observations, we expect that the TDDW effect will be more important for characterizing the Lamb-Dicke parameters
with higher accuracy in longer ion chains. 

Note that here we assumed a fixed physical distance between neighboring ions.
Thus, as $N$ increases, the spacing between the mode frequencies decreases,
which leads to more severe cross-mode coupling for a fixed qubit-state Rabi frequency. 
The effects of the mode-frequency spacing in the accuracy of $\eta_{j,k}$ estimations 
is discussed in Appendix~\ref{app:spacing}.

\subsection{Sign problem}
\label{subsec:sign}

The sign of $\eta_{j,k}$ relative to other Lamb-Dicke parameters determines the gate-pulse design on many trapped-ion quantum computers~\cite{Wu18, Grzesiak20, Bentley20, Kang21}, hence directly affecting the quantum-computational fidelity.
Unfortunately, conventional mode-characterization methods cannot distinguish the sign of $\eta_{j,k}$
because the qubit population is independent of the sign in the baseline model [see (\ref{eq:Pt_baseline})].
Here, we show that the sign of $\eta_{j,k}$ can be distinguished 
using the NN model (\textit{Model 4}).

To start, in order to distinguish the sign of $\eta_{j,k}$ using BSB transitions,
we need to consider more than one ion, 
as the sign of $\eta_{j,k}$ is well-defined only when
the \textit{relative} motion between different ions is described.
Also note that with a single mode, for different signs of $\eta_{j,k}$,
ions move in different relative directions,
but the qubit populations undergo exactly the same evolution. 
Only when we consider at least two ions and two modes simultaneously,
the sign of $\eta_{j,k}$ determines whether the symmetry of two ions' participation in one mode 
is the same or the opposite from that in the other mode, 
a difference that affects the qubit populations.

By driving two ions to couple to two different modes in parallel via illuminating the two ions with the same two-tone beam,
where each tone is resonant to the respective mode frequency,
the corresponding BSB transitions to the two modes simultaneously occur on the two ions. 
The predicted evolutions, one with the same symmetry for both modes and the other with the opposite symmetries between the two modes, become drastically different from each other.
This enables us to determine which symmetry, hence the sign of $\eta_{j,k}$, is the correct one, directly from the signal generated by the experiment.

\begin{figure}[ht!]
\includegraphics[width=8.6cm]{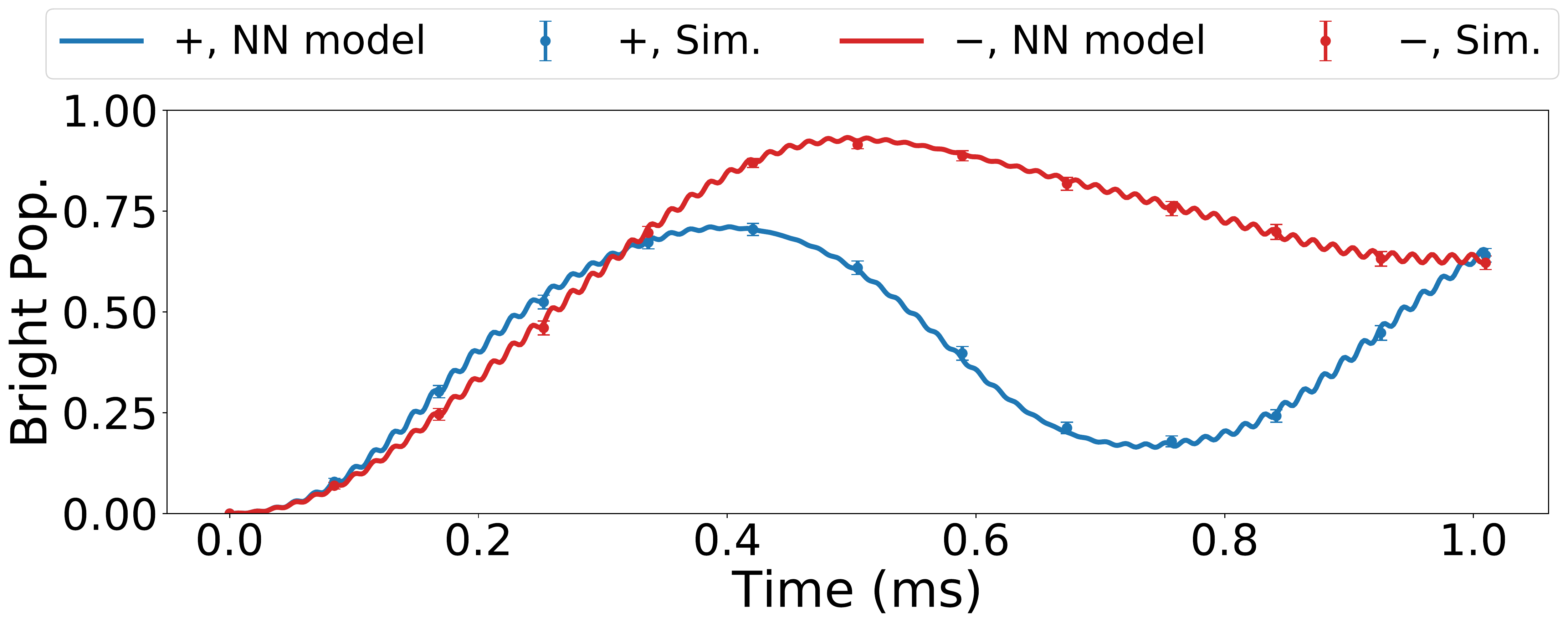}
\caption{Distinguishing the sign of $\eta_{j,k}$. 
The first and second ions of a five-ion chain are both driven with two tones, 
each tone with laser coupling frequency \mbox{$\omega_1^{\rm{qbt}} + \omega_1$}
\mbox{($\omega_2^{\rm{qbt}} + \omega_2$)} and
qubit-state Rabi frequencies \mbox{$\Omega_1 = 2\pi \times $ 30 kHz}
(\mbox{$\Omega_2 = 2\pi \times $ 9 kHz}). 
The bright-state populations of the first ion qubit
when $\eta_{1,1} = 0.0119$ (blue) and $-0.0119$ (red)
are predicted by the NN model (Eq.~(\ref{eq:HNN}), lines) with \mbox{$J = K = \{1, 2\}$},
and simulated using the entire BSB Hamiltonian (Eq.~(\ref{eq:HI}), dots). 
The entire set of mode parameters used here can be found in Appendix~\ref{app:values}.
Error bars show the shot noise for 1000 shots.
}
\label{fig:sign}
\end{figure}

Figure~\ref{fig:sign} shows an example of the different evolutions predicted, where we vary the sign of $\eta_{1,1} = \pm 0.0119$ with respect to predetermined values of 
$\eta_{1,2} = 0.0335$, $\eta_{2,1} = -0.0521$, and $\eta_{2,2} = -0.0705$ for $N = 5$. 
Both the first and second ions are driven with two tones,
which are resonant to the first and second mode frequencies $\omega_1$ and $\omega_2$, respectively.
The first tone was driven with the qubit-state Rabi frequency of \mbox{$2\pi \times 30$ kHz}
and the second tone was driven with \mbox{$2\pi \times 9$ kHz}, so as to roughly match
the resulting Rabi frequency for the transition between $\left|0,0\right>_{1,1}$ and  $\left|1,1\right>_{1,1}$
and that between $\left|0,0\right>_{1,2}$ and  $\left|1,1\right>_{1,2}$.

As expected, the population curves when \mbox{$\eta_{1,1} = \pm 0.0119$} are clearly distinguishable,
and are accurately predicted by the NN model.
This shows that the sign of $\eta_{j,k}$ can reliably be distinguished by 
inducing all four possible BSB transitions between two ions and two modes simultaneously, when carefully choosing parameters
and comparing the observed evolution with that predicted by the NN model.

\subsection{Characterization-experiment time}
\label{subsec:exptime}

The characterization-experiment time of the basic \{improved\} protocol,
given by (\ref{eq:exptimebl}) \{(\ref{eq:exptimeim})\}, depends on the following parameters: 
(i) $M^{(0)}_\Delta$ \{$M_\Delta$\}, the number of detunings scanned in the frequency scan,
(ii) $S^{(0)}$ \{$S_\Delta$ and $S_t$\}, the number of shots,
and (iii) $\tilde{\tau}^{(0)}$ \{$\tilde{\tau}_\Delta$ and $\tilde{\tau}_i$\}, the cycle time. 
Our goal is to minimize (i)-(iii), whenever applicable, while delivering a pre-determined target accuracy in estimating $\eta_{j,k}$.
Note, achieving the target accuracy is primarily hindered by the shot noise
and the uncertainties in other parameters, such as $\omega_k$.

To be consistent with Sec.~\ref{subsec:Accuracy}, we fix $M_t = 20$, and
\mbox{$\tau_{M_t} = 2.5 \sqrt{N} (\Omega_0 |\langle \vec{K} \rangle| /\sqrt{2m \langle \omega_{\rm mode} \rangle})^{-1}$},
where $N=5$ here.
Also, to compare the total experiment times on an equal footing,
we set $S_\Delta = M_t S_t$ for the improved protocol
and compare the value with $S^{(0)}$ of the basic protocol,
which uses $\tau^{(0)} = \tau_{M_t}/2$. 
Therefore, the knobs we can turn are $\Omega_0$, $S^{(0)}$, and $M_\Delta^{(0)}$
\{$\Omega_0$, $S_t$, $M_\Delta$, and $\tau_\Delta$\} for the basic \{improved\} protocol.

First, we find the number of shots $S^{(0)}$ \{$S_t$\} of the basic \{improved\} protocol
required to reach a small uncertainty in $\eta_{j,k}$. 
Here, we fit the simulated qubit populations, 
with uncertainties given by the photon and phonon shot noise combined, using \textit{Model 2}.
In order to isolate the effects of the shot noise, we assume
perfect knowledge of the mode frequencies $\omega_k$.
We use \mbox{$\Omega_0 = 2\pi \times 10$ kHz},
although the effect of shot noise is not significantly affected by $\Omega_0$.

\begin{figure}[ht!]
\includegraphics[width=8.6cm]{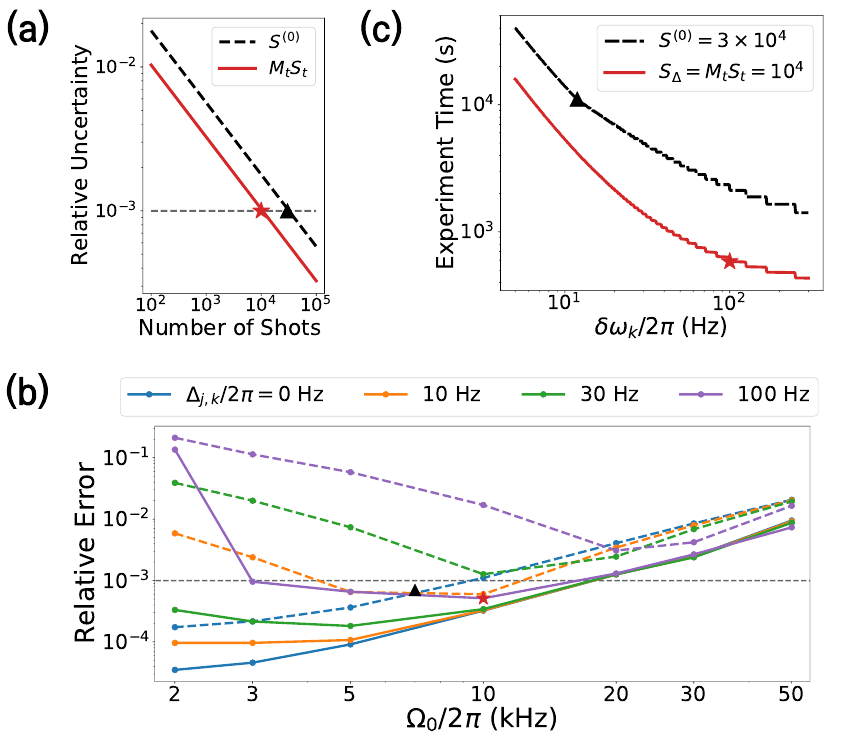}
\caption{
(a) Relative uncertainties of $\eta_{j,k}$ for various numbers of shots
of the basic protocol $S^{(0)}$ (black, dashed)
and the improved protocol $M_t S_t$ (red, solid). 
(b) Relative errors of $\eta_{j,k}$ for various qubit-state Rabi frequencies $\Omega_0$
and detunings $\Delta_{j,k}$,
measured by the basic (dashed) and improved (solid) protocols. 
(c) Characterization-experiment times of 
the basic protocol $T^{(0)}$ (black, dashed) and the improved protocol $T$ (red, solid), 
for various maximum allowed uncertainties in the mode frequencies $\delta \omega_k$
(see Appendix~\ref{app:chartime} for details).
$N=5$, and errors and uncertainties are averaged over all $N^2$ values of $\eta_{j,k}$,
excluding the Lamb-Dicke parameters corresponding to a node ($\eta_{j,k} < 10^{-4}$).
The entire set of mode parameters used here can be found in Appendix~\ref{app:values}.
The markers $\blacktriangle$ \{{\color{red} $\star$}\}
correspond to the parameters used in the basic \{improved\} protocol,
shown in Table~\ref{tab:protocol}. 
}
\label{fig:exptime}
\end{figure}

Figure~\ref{fig:exptime}(a) shows the mean relative uncertainty for various values of $S^{(0)}$ and $M_t S_t$.
The uncertainty is proportional to the inverse of square root of the number of shots. 
When $S^{(0)} = M_t S_t$,
the improved protocol always achieves a smaller uncertainty in $\eta_{j,k}$ than the basic protocol.
As explained in Sec.~\ref{sec:Protocols},
the improved protocol measures the entire $\bar{P}_{j,k}(t)$ curve,
which includes points where the qubit populations are maximally sensitive to the value of $\eta_{j,k}$.
This allows for a smaller uncertainty in $\eta_{j,k}$, compared to that obtained by the basic protocol,
as measurements conducted at a fixed timestamp $\tau^{(0)}$
cannot make all populations of $N$ qubits sensitive to $\eta_{j,k}$.
In particular, to reach an average uncertainty below $10^{-3}$, 
the basic \{improved\} protocol requires $S^{(0)} = 3 \times 10^4$ \{$M_t S_t = 10^4$\},
marked as $\blacktriangle$ \{{\color{red} $\star$}\}.

Next, we aim to find the qubit-state Rabi frequency $\Omega_0$, 
which determines the BSB-transition time $\tau^{(0)}$ \{$\tau_i$\},
and the frequency-scan parameters $M_\Delta^{(0)}$ \{$M_\Delta$ and $\tau_\Delta$\}
of the basic \{improved\} protocol, required to measure $\eta_{j,k}$ to within a target accuracy. 
To do so, we fit the qubit populations,
simulated with various values of $\Omega_0$ and detunings $\Delta_{j,k}$,
once again using \textit{Model 2}, but this time without assuming knowledge of $\Delta_{j,k}$. 
Here, we use the number of shots marked in Fig.~\ref{fig:exptime}(a),
but assume that the measured qubit probabilities are correct without shot noise,
in order to isolate the effects of the cross-mode coupling
and the inaccuracy of mode-frequency estimation.

Now, the qubit population, to be used for our fit, depends on $|\Delta_{j,k} / \Omega_0|^2$ up to the leading order 
[see (\ref{eq:Pt_baseline})]. 
Hence, the error due to nonzero $\Delta_{j,k}$ decreases as $\Omega_0$ increases. 
However, when $\Omega_0$ becomes too large, the error due to the cross-mode coupling becomes dominant, resulting in the tug of war.
From this tug of war, in principle, we can find the optimum set of parameters $\Omega_0$ and $\delta \omega_k$
that allows the measurement of $\eta_{j,k}$ with a prescribed target accuracy,
where $\delta \omega_k$ is the upper bound of $|\Delta_{j,k}|$, 
or the maximum allowed uncertainty in the mode frequencies.

There is one important subtlety to note here for the choice of $\delta \omega_k$:
While it would be ideal to admit arbitrarily small values of $\delta \omega_k$ to constrain the mode frequency uncertainties, a care needs to be taken to ensure shot noise does not drown out the population signal we aim to measure. For example, we need the difference between the qubit population
measured at $\mu_j = \omega_k$ and that measured at $\mu_j = \omega_k + \delta \omega_k$ to exceed the lower bound of the sum of shot noise, to unambiguously tell at which $\mu_j$ the population peaks.
Recall that for a given $\Omega_0$,
the frequency-scan evolution time is fixed to $\tau^{(0)}$ for the basic protocol. This, together with the fixed number of shots that were determined earlier, then results in the minimal possible $\delta \omega_k$ for the basic protocol. See Appendix~\ref{app:chartime} for details. 

We note in passing that the improved protocol does not necessarily suffer from such minimal-$\delta \omega_k$ constraint. This is so, since $\tau_\Delta$, unlike $\tau^{(0)}$, is not yet determined for the improved protocol, and we can thus absorb the shot-noise constraint into our choice of $\tau_\Delta$ itself. 
As we show later, we judiciously leverage this degree of freedom to enable much shorter characterization-experiment time, which -- we remind the readers -- is the goal of this section.

Once $\delta \omega_k$ are determined for both the basic and improved protocols, the number of detunings $M_\Delta^{(0)}$ and $M_\Delta$ can be computed. In particular, for respectively determined $\delta \omega_k$ we have
$\lceil \delta \omega_{k, \text{prior}} / 2\delta \omega_k \rceil$ as the number of detunings,
where \mbox{$\delta \omega_{k, \text{prior}}$} is the 
width of range of possible values for each $\omega_k$ that we assume to be given as a prior.

Figure~\ref{fig:exptime}(b) shows the mean relative errors in estimating $\eta_{j,k}$
as a function of $\Omega_0$. Considered are multiple $\Delta_{j,k}$ values.
Using this figure, when provided with a pre-determined target accuracy in $\eta_{j,k}$ measurement,
we can determine the values of $\Omega_0$ and $\delta \omega_k$ that 
will meet the target accuracy.
For example, if we want the relative uncertainty to be lower than $10^{-3}$,
a reasonable choice for the basic \{improved\} protocol would be
\mbox{$\Omega_0/2\pi = $ 7 \{10\} kHz} 
and \mbox{$\delta \omega_k / 2\pi = $ 12 \{100\} Hz},
marked as $\blacktriangle$ \{{\color{red} $\star$}\}.
As explained in Sec.~\ref{sec:Protocols},
the improved protocol fits the entire $\bar{P}_{j,k}(t)$ curve,
which has distinguishable effects from varying $\eta_{j,k}$ and $\Delta_{j,k}$,
allowing for a more accurate measurement in the presence of larger detuning,
compared to the basic protocol that fits the population at a single timestamp. 
The chosen value of $\delta \omega_k$ for the improved protocol leads to \mbox{$\tau_\Delta = 0.57$ ms},
as explained above and in Appendix~\ref{app:chartime}. 
Also, $\delta \omega_k$ for the basic \{improved\} protocol gives
\mbox{$M_\Delta^{(0)} = 43$} \mbox{\{$M_\Delta = 5$\}}, 
where we assumed the width of prior \mbox{$\delta \omega_{k, \text{prior}} = 2\pi \times 1$ kHz}. 

Now, with all the parameters of the protocols determined, 
we compare the characterization-experiment times of the basic and improved protocols
given in (\ref{eq:exptimebl}) and (\ref{eq:exptimeim}).
As a concrete example, we assume the times for cooling, state preparation, and state detection are, respectively,
4 ms, 100 $\mu$s, and 150 $\mu$s, which are added to the BSB-transition time 
to yield the cycle time for each shot. 
Table~\ref{tab:protocol} shows the set of parameters of the two protocols.
Overall, in order to achieve the relative measurement uncertainty of the order of $10^{-3}$ 
in estimating $\eta_{j,k}$ for a five-ion chain, 
the characterization-experiment time is \mbox{$T = 586$s} for the improved protocol, 
which is about 19 times shorter than 
\mbox{$T^{(0)} = 1.11 \times 10^4$s} for the basic protocol. 
The savings of the improved protocol come from
less precision required in the frequency scan and the fewer shots used overall. 

\begin{table}[ht!]
    \centering
    \begin{tabular}{ccc|cccccc}
        $\tilde{\tau}^{(0)}$ & $S^{(0)}$ & $M^{(0)}_\Delta$ &
        $\tilde{\tau}_\Delta$ & $S_\Delta$ & $M_\Delta$ &
        $\langle \tilde{\tau}_{i} \rangle$ & $S_t$ & $M_t$ \\
        \hline
        7.86 ms & $3\times 10^4$ & 43 &
        4.82 ms & $10^4$ & 5 &
        6.90 ms & 500 & 20
    \end{tabular}
    \caption{Parameters of the basic (left) and improved (right) protocols
    that achieve the relative uncertainty in $\eta_{j,k}$ of the order of $10^{-3}$
    for a five-ion chain.
    $\langle \cdot \rangle$ is the average over $i = 1,..,M_t$.
    According to (\ref{eq:exptimebl}) and (\ref{eq:exptimeim}), 
    the characterization-experiment times of the basic and improved protocols are
    \mbox{$T^{(0)} = 1.11 \times 10^4$ s} and \mbox{$T = 586$ s}, respectively.
    }
    \label{tab:protocol}
\end{table}

Finally, to distinguish the advantage of requiring less stringent frequency-scan precision and fewer shots overall, 
Fig.~\ref{fig:exptime}(c) shows the experiment times of the two protocols
for various values of $\delta \omega_k$. 
This emphasizes that allowing larger uncertainty $\delta \omega_k$ in the mode frequencies
significantly reduces the characterization experiment time for the improved protocol.

\section{Discussion and outlook}
\label{sec:Discussion}

\subsection{Additional sources of errors}

In Sec.~\ref{subsec:exptime}, we considered multiple sources of errors in estimating the Lamb-Dicke parameters, such as shot noise, inaccurate mode frequencies, and cross-mode coupling, which altogether led to determining the characterization-experiment time for achieving a target accuracy. However, in actual experiments, there could be various other sources of errors that may affect both the accuracy and efficiency of the mode characterization. In this subsection, we briefly discuss how mode characterization needs to be performed in the presence of such realistic experimental conditions. 

Additional sources of errors, which cause discrepancies between the actual Hamiltonian and the model Hamiltonian, can be categorized into two types. First, the error source can be \textit{static} within the operation cycle (which includes calibration, characterization, and running circuits, as shown in Fig.~\ref{fig:intro}). Examples are: the offset in qubit-state Rabi frequencies, optical crosstalk, and anharmonicity of the motional modes. In such instances, the relative errors in mode-parameter estimation are lower bounded by a nonzero value. This is similar to the case of the baseline model and \textit{Model 1} in Fig.~\ref{fig:errors}, where the relative errors are lower bounded by approximately $7 \times 10^{-3}$ and $4 \times 10^{-4}$, respectively, even for the smallest Rabi frequency considered. Similar lower bounds may occur to \textit{Models 2-5} as well in the presence of effects that are not included in these models. 

Second, the error sources can be time-varying fluctuations of the physical parameters, such as the uncertainties in qubit-state Rabi frequencies, mode frequencies, and Lamb-Dicke parameters. In most cases, the average qubit population over a large number of shots approaches the value when each parameter is at its mean value of fluctuation (exceptions are: (i) the range of fluctuation in the qubit population includes zero or one, which can be avoided by carefully choosing the timestamps, and (ii) parameters drift slowly compared to the characterization-experiment time or the operation cycle, which will be discussed later). However, the standard deviations in qubit populations due to the parameter fluctuations add to those due to shot noise and contribute to the relative uncertainties in mode-parameter estimation. 

Therefore, when performing mode characterization, the target accuracies in estimating the mode parameters need to take account for the magnitude of errors of both types. Then, the protocol parameters, such as the number of shots, number of frequency-scan points, qubit-state Rabi frequency, and evolution time, can be determined from the target accuracies, such that the characterization time is minimized, similarly to the process described in Sec.~\ref{subsec:exptime} and Fig.~\ref{fig:exptime}. 

As an example, if the qubit-state Rabi frequency, with the calibration uncertainty and fluctuation combined, is within the range $[(1-\epsilon)\Omega_j, \: (1+\epsilon)\Omega_j]$, then the target inaccuracy in $\eta_{j,k}$ estimation needs to be larger than $\epsilon \eta_{j,k}$. This is because the effective Rabi frequency $\Omega^{(0)}_{j,k}$ is equal to $\eta_{j,k} \Omega_j$ up to first order in $\eta_{j,k}$. 

When the magnitude of parameter uncertainty or noise is not known, one may consider tweaking the protocol to accommodate for such a situation. For example, mode parameters may be iteratively estimated, starting from using a large Rabi frequency, then reducing it gradually at each iteration, such that the effects of cross-mode coupling are reduced. Ideally, the estimation of $\eta_{j,k}$ would converge to a value. In the presence of mode-frequency fluctuations though, too small of a Rabi frequency would result in noisy estimation due to the increased sensitivity. One suggestion may then be that to halt the reduction of Rabi frequency, as soon as the estimation via convergence can be made, before the effects of fluctuations kick in. Similarly, one may consider adaptively determining the number of shots, where the repetition of shots is halted when the estimation of a mode parameter converges, before, say, the parameter drift over the characterization-experiment time becomes significant. 

We note that our improved protocol is expected to be more robust to additional sources of errors than the basic protocol in estimating $\eta_{j,k}$ for two reasons. First, due to the separation of signals for $\omega_k$ and $\eta_{j,k}$, the improved protocol can handle a larger uncertainty in $\omega_k$ for estimating $\eta_{j,k}$ to within the same precision than the basic protocol. This allows for using a larger Rabi frequency, which leads to a reduced sensitivity to mode-frequency fluctuations. 
Second, the improved protocol requires a shorter characterization-experiment time than the basic protocol. When physical parameters drift slowly over the operation cycle, the mode parameters may deviate from the measured values after the characterization is complete. Performing a shorter characterization experiment reduces the amount of parameter drifts that occur during the operation cycle, as well as enables more frequent characterization with minimal overhead, fending off the effects of the drifts. This highlights that an efficient protocol is desirable not only for the efficiency itself but also for improved accuracy as well.

\subsection{Resource trade offs}

The problem of efficient motional-mode characterization with high accuracy
boils down to an optimization over multiple parameters that are correlated by various trade offs. 
For example, using a smaller laser power (thus smaller $\Omega_0$)
reduces the errors due to cross-mode coupling,
at the cost of requiring longer BSB-transition time and better frequency-scan precision.

There are still many degrees of freedom that can be explored beyond the scope of this paper.
For example, we fix \mbox{$\tau_{M_t} = 2.5 \sqrt{N} (\Omega_0 |\langle \vec{K} \rangle| /\sqrt{2m \langle \omega_{\rm mode} \rangle})^{-1}$}, although varying the BSB-transition time
potentially leads to interesting trade offs even at fixed $N$ and $\Omega_0$,
especially when the system is susceptible to additional sources of errors such as motional dephasing and heating~\cite{Wang20, Cetina22, Kang22}.
Also, the pulses used to probe the modes do not necessarily have a constant amplitude and drive frequency.
The trade offs for using shaped pulses that suppress errors due to
cross-mode coupling and inaccurate mode frequencies are to be published elsewhere~\cite{shapedpulse}.

The choice of protocols and models themselves can also be viewed as a part of the trade offs.
For example, a parallelized protocol reduces the complexity from $O(N^2)$ to $O(N)$,
at the cost of bringing additional considerations into the model,
such as the DW effect from the other modes being probed in parallel, which is time-dependent to be precise. 
In general, a more accurate model can be used at the cost of longer conventional-computation time. 
To exploit this trade off, a highly parallelized and efficient algorithm for the fitting routine may be explored,
performing the conventional-computation part of the protocol relatively fast,
especially relevant for long ion chains where the computation tends to slow down
(see Appendix~\ref{app:fittingroutine} for details).

Another important trade off relevant to trapped ions is 
the spacing between mode frequencies versus the physical distance between neighboring ions. 
Smaller distance between neighboring ions leads to larger spacing between the mode frequencies,
which allows smaller errors in measuring $\eta_{j,k}$ as the cross-mode-coupling effects are reduced
(see Appendix~\ref{app:spacing} for details).
This can alleviate the exponential increase of error in $N$ observed in Fig.~\ref{fig:errors}(b),
which assumes a fixed distance between neighboring ions. 
However, a smaller inter-ion distance leads to larger optical crosstalk,
as the laser beam width cannot be made arbitrarily small. 

In principle, the effects of optical crosstalk can be included in the model, which potentially enables more accurate characterization. However, this is achievable only at the cost of additional calibration devoted to measuring the magnitude of crosstalk, as well as longer conventional-computation time.

\subsection{Outlook}
Even after analyzing every existing trade off and finding the optimally efficient protocol,
the characterization-experiment time for longer ion chains 
can still take a significant portion of a typical trapped-ion system's operation cycle.
To save the experiment time as much as possible, clever calibration techniques,
such as using Bayesian inference \cite{Gerster21}, can be combined with this work. 

Overall, we developed an efficient method of characterizing the motional modes with high accuracy
and analyzed the required resources. 
Such system characterization will be a crucial component of a 
scalable, fault-tolerant trapped-ion quantum computer. 
We hope this paper motivates developing more efficient characterization schemes,
not only for trapped ions but also for various other platforms of quantum computing. 


\section*{Acknowledgements}
M.K. and Q.L. thank Kenneth R. Brown for helpful discussions.
This material was based on work supported by IonQ Inc., while Q.L., M.L., and Y.N. were working at IonQ Inc.
Any opinion, finding, and conclusions or recommendations expressed in this material
are those of the authors and do not necessarily reflect the views of IonQ Inc. 


\bibliography{bib}

\appendix

\section{Hamiltonian simulation}
\label{app:simulation}

Here we outline the details of simulating the evolution 
with respect to the BSB Hamiltonian $\hat{H}_{I}$ in (\ref{eq:HI}). 
The Hamiltonian in the NN model $\hat{H}_{NN}$ in (\ref{eq:HNN}) can also be simulated equivalently.

Given the vector of phonon numbers $\vec{n}$, 
where its $k$-th component is the initial phonon number of the $k$-th mode, 
each mode is approximated to a four-level system, 
where the lowest level represents the Fock state $|\text{max}(n_{k}-1, 0)\rangle$.
Since the BSB transition primarily occurs between $|n_k\rangle$ and $|n_k + 1\rangle$,
this four-level approximation is sufficiently accurate. 
Therefore, the Hamiltonian of $N$ qubits and $N'$ modes is simulated in a Hilbert space of dimension $2^N \times 4^{N'}$.

To simulate the evolution, we first divide the time evolution
into fine sub-steps of length $2\pi \times 0.002 / \Omega_0$. 
At each sub-step, we fill in all entries of the Hamiltonian that correspond to a blue-sideband transition.
Here, we use the rotating-wave approximation 
\mbox{$\tilde{\omega}_{j} - \omega_{j}^{\rm{qbt}} \gg \eta_{j,k}\Omega_0$}
to ignore the carrier and red-sideband transitions,
which is valid for the range of $\Omega_0$ and the errors considered in this paper.
Note that up to this stage we do not expand any exponential in the Hamiltonian to a Taylor series.

Now, we integrate each entry over time, from current to next sub-step. 
The resulting matrix is the first term of the Magnus series, and with sufficiently short sub-steps,
the exponentiation of $-i$ times this matrix approximates the unitary evolution
with respect to the Hamiltonian with sufficiently high accuracy. 
Then, the exponentiation is evaluated by expanding the Taylor series up to the fifth order. 
The resulting matrix is multiplied to the state vector at the current sub-step,
which yields the state vector at the next sub-step. 
This is repeated for all sub-steps until we obtain the state vector at the end of the evolution.
Finally, the state vector is projected to the subspace of $N$ qubits to give the qubit populations.

For the parameters such as the number of levels for each mode, 
the length of sub-step, and the number of terms in the Taylor series,
we perform convergence tests and verify that the errors in the qubit populations are sufficiently low
with the parameters used in the simulations.

\section{Algorithm for fitting routine}
\label{app:fittingroutine}

In \textit{Models 1-5}, the qubit population $\bar{P}_{j,k}(t)$
depend not only on $\eta_{j,k}$, but also on other Lamb-Dicke parameters $\eta_{j',k'}$ ($(j',k') \neq (j,k)$).
Naively fitting the populations $\bar{P}_{j',k'}(t)$ ($j=1,..,N$, $k=1,..,N'$) altogether,
with $N \times N'$ Lamb-Dicke parameters $\eta_{j',k'}$ as fitting parameters,
requires impractically long conventional-computation time. 

To avoid this issue, we employ a fitting routine of multiple iterations, shown in Algorithm~\ref{algo:routine}. 
In the improved protocol's case, we only use two fitting parameters $\eta_{j,k}$ and $\Delta_{j,k}$
when fitting the set of qubit populations $\bar{P}_{j,k}(\tau_i)$ ($i = 1,..,M_t$).
In the first iteration, the initial-guess values $\eta_{j',k'}^{(0)}$
are used for evaluating $\bar{P}_{j,k}(\tau_i)$ with \mbox{\textit{Models 1-5}},
where the superscript $^{(0)}$ here represents initial guess.
The fitted Lamb-Dicke parameter for each $j,k$ is stored as $\eta_{j,k}^{(1)}$.
In the $r$-th iteration ($r \geq 2$), the Lamb-Dicke parameters obtained from the previous iteration
$\eta_{j',k'}^{(r-1)}$ are used to obtain $\eta_{j,k}^{(r)}$.
The iterations are performed until the Lamb-Dicke parameters obtained in consecutive iterations converge. 
For the basic protocol, we use an equivalent algorithm 
with the set of qubit populations $\bar{P}_{j,k}(\tau_i)$ ($i = 1,..,M_t$)
replaced by a single qubit population $\bar{P}_{j,k}(\tau^{(0)})$ for each $j,k$.
Typically, with reasonably good initial-guess values, repeating two rounds is sufficient. 
This significantly reduces the computation time of the fitting for large $N$. 

\begin{algorithm}
\caption{Fitting routine for improved protocol} \label{algo:routine}
\begin{algorithmic}
\Require $\bar{P}_{j,k}(\tau_i)$, $\eta^{(0)}_{j,k}$, $\epsilon$, \textit{Model}
\State $r \gets 0$
\While{$|\eta^{(r)}_{j,k} - \eta^{(r-1)}_{j,k}| > \epsilon$ for any $j,k$}
\State $r \gets r+1$
\For{$j=1,..,N$, $k=1,..,N'$, $\forall\: \vec{n}$}
    \State Evaluate $D_{j,k}(n_k)$ and $P^{(\vec{n})}_{j,k}(t)$ at all time steps
    \State \quad with $\eta_{j,k} = \eta^{(r-1)}_{j,k}$
\EndFor
\For{$j=1,..,N$, $k=1,..,N'$}
    \State Fit $\eta_{j,k}$, $\Delta_{j,k}$ into set of $\bar{P}_{j,k}(\tau_i)$~'s ($i = 1,..,M_t$),
    \State \quad using the \textit{Model} 
    \State \quad with $\eta_{j',k'} = \eta^{(r-1)}_{j',k'}$ $\forall (j',k') \neq (j,k)$
    \State $\eta^{(r)}_{j,k} \gets \eta_{j,k}$
\EndFor
\EndWhile
\State \Return $\eta^{(r)}_{j,k}$
\end{algorithmic}
\end{algorithm}

All loops over the ion indices $j$ and mode indices $k$ are embarrassingly parallel. 
Evaluating $\bar{P}_{j,k}(\tau_i)$ from the \textit{Model},
which requires evaluating $P^{(\vec{n})}_{j,k}(\tau_i)$ for all $\vec{n}$, is also parallelizable. 
With $N \times N'$ computing nodes, each equipped with number of cores equal to the number of $\vec{n}$~'s, 
the computation time of each fitting does not necessarily increase with $N$. 
This allows the computational part of the mode characterization to be scalable for long ion chains,
even when using models where the qubit populations are correlated to all modes.

\section{Mode-frequency estimation}
\label{app:chartime}

Here we outline the details of the calculations in Sec.~\ref{subsec:exptime}. 
In particular, we show how the uncertainty in the mode-frequency estimation $\delta \omega_k$
is related to the BSB-transition times $\tau^{(0)}$ and $\tau_\Delta$
used in the frequency scan of the basic and improved protocols.
For the number of shots, we use the value marked in Fig.~\ref{fig:exptime}(a),
which is \mbox{$S^{(0)} = 3 \times 10^4$} \{\mbox{$S_\Delta = 10^4$}\}
for the basic \{improved\} protocol.

Let us consider the qubit population undergoing BSB transition as a function of both time $t$ and detuning $\Delta$ in the baseline model [see (\ref{eq:Pt_baseline})], given by
\begin{equation}
    \bar{P}^{(0)}(t, \Delta) = \frac{[\Omega^{(0)}]^2}{[\Omega^{(0)}]^2 + \frac{\Delta^2}{4}} 
    \sin^2 \left(\sqrt{[\Omega^{(0)}]^2 + \frac{\Delta^2}{4}} t \right),
\end{equation}
where the ion and mode indices $j,k$ are omitted. 

For the mode frequency to be distinguished with uncertainty less than $\delta \omega$,
the difference between the qubit populations measured at detunings 0 and $\delta \omega$
should exceed the sum of shot noise. 
Using that the sum of shot noise is minimized when $\bar{P}^{(0)}(\tau, 0) = 1$,
the necessary condition of measuring the mode frequency up to uncertainty $\delta \omega$
with BSB-transition time $\tau$ and number of shots $S$ becomes
\begin{equation}
    \left|\bar{P}^{(0)}(\tau, 0) - \bar{P}^{(0)}(\tau, \delta \omega)\right| 
    \geq \sqrt{\frac{\bar{P}^{(0)}(\tau, \delta \omega) \big(1-\bar{P}^{(0)}(\tau, \delta \omega) \big)}{S}}, \nonumber
\end{equation}
which leads to
\begin{equation}\label{eq:deltaomega}
    \delta \omega \geq \frac{2}{\tau}\left(
    \Big(\pi - \sin^{-1}\sqrt{\frac{S}{1+S}}\Big)^2 - \Big(\frac{\pi}{2}\Big)^2 \right)^{1/2}.
\end{equation}

For the basic protocol, \mbox{$\Omega_0 = 2\pi \times 7$ kHz} as marked in Fig.~\ref{fig:exptime}(b) gives
$\tau^{(0)} = 2.5 \sqrt{N} \times (2\Omega_0 |\langle \vec{K} \rangle| /\sqrt{2m \langle \omega_{\rm mode} \rangle}|)^{-1} = 3.61$ ms.
Then, for $\tau = \tau^{(0)}$ and $S = S^{(0)}$, (\ref{eq:deltaomega}) gives \mbox{$\delta \omega \geq 2\pi \times 12$ Hz}.
The lower bound roughly agrees with Fig.~\ref{fig:exptime}(b), 
where the maximum detuning that allows
the mean relative error in $\eta_{j,k}$ to be lower than $10^{-3}$ is
approximately \mbox{$2\pi \times 10$ Hz}.
Therefore, we use \mbox{$\delta \omega_k = 2\pi \times 12$ Hz} for the basic protocol
in calculating the parameters of Table~\ref{tab:protocol}. 

Figure~\ref{fig:exptime}(c) plots the experiment times for various values of $\delta \omega_k$.
Here we use the lower bounds of the BSB-transition times, given by
\begin{align}
    \tau^{(0)} = \min \Bigg[ &3.61 \text{ ms}, \nonumber\\
    &\frac{2}{\delta \omega_k}\left(
    \Big(\pi - \sin^{-1}\sqrt{\frac{S^{(0)}}{1+S^{(0)}}}\Big)^2 - \Big(\frac{\pi}{2}\Big)^2 \right)^{1/2}
    \Bigg] \nonumber
\end{align}
and
\begin{align}
    \tau_\Delta = \frac{2}{\delta \omega_k} \left(
    \Big(\pi - \sin^{-1}\sqrt{\frac{S_\Delta}{1+S_\Delta}}\Big)^2 - \Big(\frac{\pi}{2}\Big)^2 \right)^{1/2}. \nonumber
\end{align}
In particular, for the improved protocol,
\mbox{$\delta \omega_k = 2\pi \times 100$ Hz} as marked in Fig.~\ref{fig:exptime}(b)
yields \mbox{$\tau_\Delta = 0.57$ ms}, which is the value used in Table~\ref{tab:protocol}. 

Note that in the improved protocol, we fix \mbox{$S_\Delta = M_t S_t = 10^4$}
for a fair comparison with the basic protocol.
In practice, $S_\Delta$ can be set as smaller than $M_t S_t$, 
which further reduces the frequency-scan experiment time.

\section{Mode-frequency spacing}
\label{app:spacing}

The error in estimating $\eta_{j,k}$ due to the cross-mode coupling can be reduced by
using a smaller qubit-state Rabi frequency $\Omega_0$,
but this increases the evolution time in order to fix the degree of BSB transition. 
An alternative way of reducing the effect of the cross-mode coupling is
to increase the spacing between the mode frequencies. 
This can be achieved by reducing the physical distance between neighboring ions.

\begin{figure}[ht]
\includegraphics[width=8.6cm]{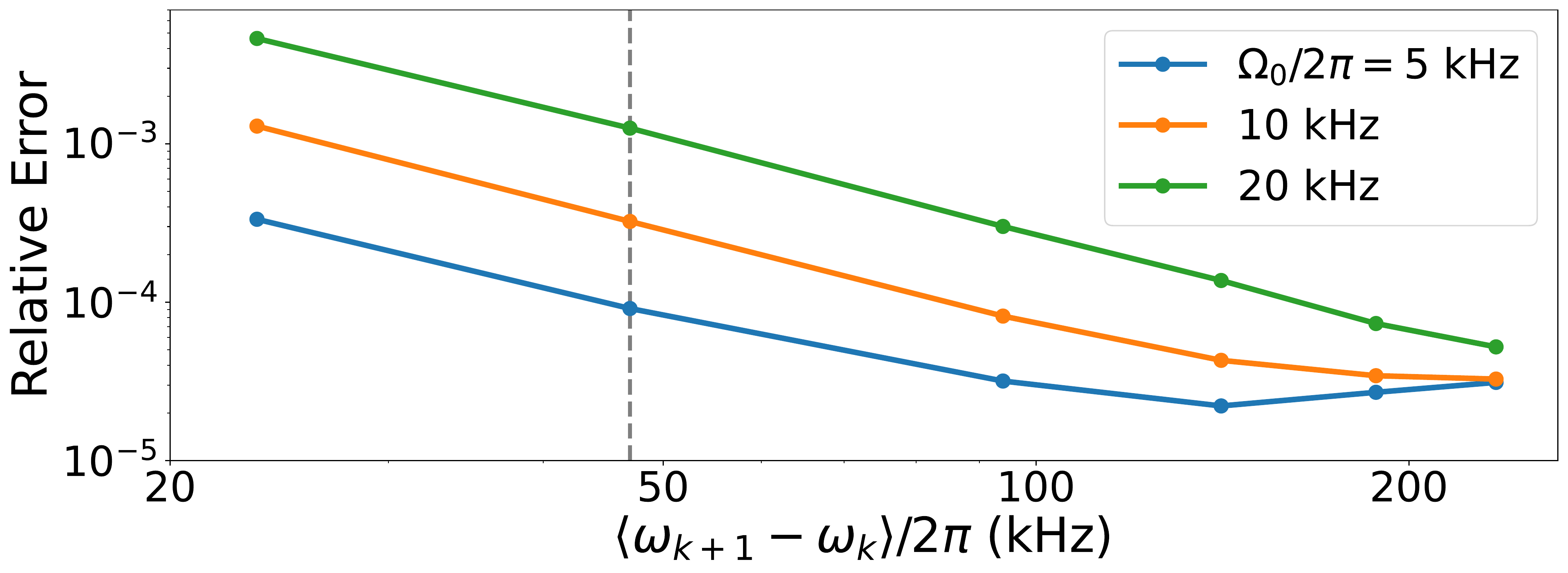}
\caption{
Relative errors of $\eta_{j,k}$ measured using \textit{Model 2}
for various qubit-state Rabi frequencies $\Omega_0$
and average spacings between neighboring mode frequencies \mbox{$\langle \omega_{k+1} - \omega_k \rangle$}.
Here we use \mbox{$N = N' = 5$}.
Excluding the Lamb-Dicke parameters corresponding to a node \mbox{($\eta_{j,k} < 10^{-4}$)},
the errors are averaged over all $N^2$ values of $\eta_{j,k}$.
The entire set of mode parameters used here can be found in Appendix~\ref{app:values}.
Dashed line indicates the average spacing used in the main text.
}
\label{fig:spacing}
\end{figure}

Figure~\ref{fig:spacing} shows the mean errors of $\eta_{j,k}$ obtained
by fitting the set of populations $\bar{P}_{j,k}(\tau_i)$ \mbox{$(i = 1,..,M_t)$} into \textit{Model 2},
for various values of average spacing between neighboring mode frequencies $\langle \omega_{k+1} - \omega_k \rangle$,
where the average is over $k = 1,..,N'-1$.
For errors larger than $3 \times 10^{-5}$,
as the average spacing increases,
the error decreases as a power law of \mbox{$(\langle \omega_{k+1} - \omega_k \rangle)^{-2}$}.
Smaller error cannot be achieved by increasing the mode-frequency spacings, as there exists errors due to effects other than the cross-mode coupling that are not captured in \mbox{\textit{Model 2}}, 
such as the time dependence of the DW effect. 

Reducing the physical distance between neighboring ions leads to larger spacing between the mode frequencies,
and therefore allows accurate characterization of the mode parameters
with larger $\Omega_0$ and shorter experiment time.
We note that for individually addressed operations,
laser beamwidth sets a lower bound on the physical distance between ions.

\section{Values of mode parameters}
\label{app:values}

Here we present the values of the mode frequencies $\omega_k$ and the Lamb-Dicke parameters $\eta_{j,k}$ used in the simulations. The mode parameters are obtained by numerically solving the normal modes of equidistantly spaced ions trapped by a modelled potential of an HOA2.0 trap~\cite{HOA}. In all simulations we use $N'=N$, which corresponds to a typical laser alignment.

\begin{table*}[ht]
\renewcommand*{\arraystretch}{1.5}
\begin{tabular}{ c | c | c | c}
\hline
& $k = 1$ & $k = 2$ & $k = 3$\\
\hline
\begin{tabular}{@{}c@{}}$\omega_k / 2\pi$ \\ (MHz)\end{tabular} & 
2.9574 & 3.0542 & 3.1222 \\
\hline
$\eta_{j=1,k}$ & $-0.0457$ & $0.0776$ & $0.0625$\\
$\eta_{j=2,k}$ & $0.0909$ & $-2.77 \times 10^{-6}$ & $0.0629$ \\
$\eta_{j=3,k}$ & $-0.0457$ & $-0.0776$ & $0.0625$ \\
\hline
\end{tabular}
\caption{Values of mode parameters for $N = N' = 3$.}\label{tab:valuesN3}
\end{table*}

\begin{table*}[ht]
\renewcommand*{\arraystretch}{1.5}
\begin{tabular}{ c | c | c | c | c}
\hline
& $k = 1$ & $k = 2$ & $k = 3$ & $k = 4$\\
\hline
\begin{tabular}{@{}c@{}}$\omega_k / 2\pi$ \\ (MHz)\end{tabular} & 
2.9467 & 3.0263 & 3.0894 & 3.1333 \\
\hline
$\eta_{j=1,k}$ & $0.0239$ & $0.0551$ & $0.0735$ & $0.0542$\\
$\eta_{j=2,k}$ & $-0.0753$ & $-0.0552$ & $0.0234$ & $0.0541$ \\
$\eta_{j=3,k}$ & $0.0753$ & $-0.0552$ & $-0.0234$ & $0.0541$\\
$\eta_{j=4,k}$ & $-0.0239$ & $0.0551$ & $-0.0735$ & $0.0542$\\
\hline
\end{tabular}
\caption{Values of mode parameters for $N = N' = 4$.}\label{tab:valuesN4}
\end{table*}

\begin{table*}[ht]
\renewcommand*{\arraystretch}{1.5}
\begin{tabular}{ c | c | c | c | c | c}
\hline
& $k = 1$ & $k = 2$ & $k = 3$ & $k = 4$ & $k = 5$\\
\hline
\begin{tabular}{@{}c@{}}$\omega_k / 2\pi$ \\ (MHz)\end{tabular} & 
2.9526 & 3.0155 & 3.0687 & 3.1115 & 3.1407 \\
\hline
$\eta_{j=1,k}$ & $0.0119$ & $0.0335$ & $-0.0586$ & $0.0694$ & $0.0486$\\
$\eta_{j=2,k}$ & $-0.0526$ & $-0.0705$ & $0.0307$ & $0.0330$ & $0.0482$ \\
$\eta_{j=3,k}$ & $0.0814$ & $1.66 \times 10^{-5}$ & $0.0569$ & $1.12 \times 10^{-5}$ & $0.0481$\\
$\eta_{j=4,k}$ & $-0.0526$ & $0.0705$ & $0.0307$ & $-0.0330$ & $0.0483$\\
$\eta_{j=5,k}$ & $0.0119$ & $-0.0335$ & $-0.0586$ & $-0.0694$ & $0.0487$\\
\hline
\end{tabular}
\caption{Values of mode parameters for $N = N' = 5$.}\label{tab:valuesN5}
\end{table*}

\begin{table*}[ht]
\renewcommand*{\arraystretch}{1.5}
\begin{tabular}{ c | c | c | c | c | c | c}
\hline
& $k = 1$ & $k = 2$ & $k = 3$ & $k = 4$ & $k = 5$ & $k = 6$\\
\hline
\begin{tabular}{@{}c@{}}$\omega_k / 2\pi$ \\ (MHz)\end{tabular} & 
2.9444 & 2.9983 & 3.0465 & 3.0881 & 3.1225 & 3.1443 \\
\hline
$\eta_{j=1,k}$ & $-0.00603$ & $-0.0192$ & $0.0395$ & $-0.0592$ & $-0.0658$ & $0.0453$\\
$\eta_{j=2,k}$ & $0.0338$ & $0.0624$ & $-0.0591$ & $0.0146$ & $-0.0375$ & $0.0439$ \\
$\eta_{j=3,k}$ & $-0.0711$ & $-0.0433$ & $-0.0314$ & $0.0473$ & $-0.0123$ & $0.0432$\\
$\eta_{j=4,k}$ & $0.0711$ & $-0.0432$ & $0.0314$ & $0.0473$ & $0.0122$ & $0.0432$\\
$\eta_{j=5,k}$ & $-0.0338$ & $0.0624$ & $0.0591$ & $0.0146$ & $0.0375$ & $0.0439$\\
$\eta_{j=6,k}$ & $0.00603$ & $-0.0192$ & $-0.0395$ & $-0.0592$ & $0.0658$ & $0.0453$\\
\hline
\end{tabular}
\caption{Values of mode parameters for $N = N' = 6$.}\label{tab:valuesN6}
\end{table*}

\begin{table*}[ht]
\renewcommand*{\arraystretch}{1.5}
\begin{tabular}{ c | c | c | c | c | c | c | c}
\hline
& $k = 1$ & $k = 2$ & $k = 3$ & $k = 4$ & $k = 5$ & $k = 6$ & $k=7$\\
\hline
\begin{tabular}{@{}c@{}}$\omega_k / 2\pi$ \\ (MHz)\end{tabular} & 
2.9444 & 2.9880 & 3.0295 & 3.0679 & 3.1020 & 3.1312 & 3.1471 \\
\hline
$\eta_{j=1,k}$ & $-0.00341$ & $-0.0113$ & $-0.0253$ & $-0.0431$ & $-0.0577$ & $0.0627$ & $0.0440$\\
$\eta_{j=2,k}$ & $0.0216$ & $0.0478$ & $0.0623$ & $0.0466$ & $0.00367$ & $0.0396$ & $0.0407$ \\
$\eta_{j=3,k}$ & $-0.0552$ & $-0.0612$ & $-0.0100$ & $0.0443$ & $0.0375$ & $0.0193$ & $0.0389$\\
$\eta_{j=4,k}$ & $0.0738$ & $-9.10 \times 10^{-5}$ & $-0.0538$ & $5.40 \times 10^{-5}$ & $0.0486$ & $1.10 \times 10^{-5}$ & $0.0383$\\
$\eta_{j=5,k}$ & $-0.0551$ & $0.0612$ & $-0.0102$ & $-0.0443$ & $0.0376$ & $-0.0193$ & $0.0389$\\
$\eta_{j=6,k}$ & $0.0216$ & $-0.0477$ & $0.0623$ & $-0.0467$ & $0.00373$ & $-0.0396$ & $0.0407$\\
$\eta_{j=7,k}$ & $-0.00339$ & $0.0113$ & $-0.0252$ & $0.0430$ & $-0.0577$ & $-0.0627$ & $0.0440$\\
\hline
\end{tabular}
\caption{Values of mode parameters for $N = N' = 7$.}\label{tab:valuesN7}
\end{table*}

\end{document}